\documentstyle[art8, aaspp4,11pt]{article}

\def\etal{{\it et al.$\,\,$}}
\def\simless{\mathbin{\lower 3pt\hbox
   {$\rlap{\raise 5pt\hbox{$\char'074$}}\mathchar"7218$}}} 
\def\simgreat{\mathbin{\lower 3pt\hbox
   {$\rlap{\raise 5pt\hbox{$\char'076$}}\mathchar"7218$}}} 

\def\Rgs{{R_0}}
\def\rsun{{R_{\odot}}}
\def\msun{{M_{\odot}}}
\def\eps02{{{\epsilon_{-2}}}}

\def\porb{{P_{orb}}}

\def\kms{{km s$^{-1}$ }}

\def\vperp{{V_{\perp}}}

\def\cl{{\cal L}}

\def\thetavec{{ \mbox{\boldmath $\theta$} }}

\def\lpprime{{ L_p^{\prime} }}
\def\nfft{{N_{FFT}}}
\def\smin{{S_{min} }}
\def\vvmax{{V/V_{\rm max} }}
\def\npave{{\langle N_p \rangle }}
\def\sprime{{ S^{\prime} }}
\def\niss{{ n_{iss} }}

\def\n0hat{{\hat n_0 }}

\tighten
\singlespace
\begin{document}
\title{NEUTRON STAR POPULATION DYNAMICS.I: \\ 
MILLISECOND PULSARS}
\author{J. M. Cordes}
\affil{Astronomy Department and NAIC, Cornell University}
\author{David F. Chernoff}
\affil{Astronomy Department, Cornell University}
\centerline{In press {\it ApJ}, 482, 1997 June 20}
\begin{abstract}

We study the field millisecond pulsar population to infer its
intrinsic distribution in spin period and luminosity and to determine
its spatial distribution within the Galaxy.  Our likelihood
analysis on data from extant surveys (22 pulsars with periods $< 20$
ms) accounts for the following important selection effects: (1) the
survey sensitivity as a function of direction, spin period, and sky
coverage; (2) interstellar scintillation, which modulates the pulsed
flux and causes a net increase in search volume $\sim 30$\%; and (3)
errors in the pulsar distance scale.

Adopting power-law models (with cutoffs) for the intrinsic
distributions, the analysis yields 
a minimum period cutoff ${\rm P_{min} > 0.65}$ ms (99\% confidence), 
a period distribution $\propto P^{-2.0\pm0.33}$ and a 
pseudo-luminosity distribution $\propto L_p^{-2.0\pm0.2}$ (where $L_p
= $ flux density $\times$ distance$^2$, for $L_p \ge 1.1$ mJy
kpc$^2$).

We find that the column density of millisecond pulsars (uncorrected
for beaming effects) is $\sim 50^{+30}_{-20}$ kpc$^{-2}$ in the
vicinity of the solar system. For a Gaussian model the $z$ scale
height is $0.65^{+0.16}_{-0.12}$ kpc, corresponding to local number
density $29^{+17}_{-11}$ kpc$^{-3}$. (For an exponential model the
scale height becomes $0.50^{+0.19}_{-0.13}$ kpc and the number density
$44^{+25}_{-16}$ kpc$^{-3}$.) Estimates of the total number of MSPs
in the disk of the Galaxy and for the associated birthrate are given.
The contribution of a diffuse halo-like component (tracing the
Galactic spheroid, the halo or the globular cluster density profile)
to the local number density of MSPs is limited to $\simless 1$\% of
the midplane value.

We consider a kinematic model for the MSP spatial distribution in
which objects in the disk are kicked once at birth and then orbit in a smooth
Galactic potential, becoming dynamically well-mixed. The analysis
yields a column density $49^{+27}_{-17}$ kpc$^{-2}$ (comparable to the
above), a birth $z$ kick velocity $52^{+17}_{-11}$ km s$^{-1}$ and
a 3D velocity dispersion of $\sim84$ km s$^{-1}$. MSP velocities are
smaller than those of young, long-period pulsars by about a factor of
5.  The kinematic properties of the MSP population are discussed,
including expected transverse motions, the occurrence of asymmetric
drift, the shape of the velocity ellipsoid and the 
$z$ scale height at birth.
If MSPs are long-lived then a
significant contribution to observed MSP $z$ velocities owes to diffusive
processes that increase the scale height of old stellar populations;
our best estimate of the 1D velocity kick that is unique to MSP
evolution is $\sim 40$ km s$^{-1}$ if such diffusion is taken into account.

The scale heights of millisecond pulsars and low-mass X-ray binaries
are consistent, suggesting a common origin and that the primary
channel for forming both classes of objects imparts only low
velocities.  Binaries involving a common envelope phase and a
neutron-star forming supernova explosion can yield such objects, even
with explosion asymmetries like those needed to provide the velocity
distribution of isolated, nonspunup radio pulsars.

Future searches for MSPs may be optimized using the model results.  As
an example, we give the expected number of detectable MSPs per beam
area and the volumes of the Galaxy sampled per beam area for a
hypothetical Green Bank Telescope all sky survey.  Estimates for the
volume that must be surveyed to find a pulsar faster than $1.5$ ms are
given.  We also briefly discuss how selection effects associated with
fast binaries influence our results. 

\end{abstract}


\section{INTRODUCTION}\label{sec:intro} 

Millisecond pulsars (MSPs) differ from slower-spin pulsars in
important ways.  First, their spindown rates and derived surface
magnetic fields are several orders of magnitude smaller. MSPs have
implied fields of $10^{7.9}-10^9$ Gauss, while pulsars with periods of
order one second are characterized by magnetic fields of $10^{11} -
10^{13}$ Gauss.  Closely related is the observation that the
characteristic spin-down times of MSPs, ranging from several tenths to
tens of Gyr, far exceed those of slower-spin pulsars.  Some MSPs were
born with periods near their present-day values and are,
consequently, much younger than their spindown times (\cite{ctk94}).
However, MSPs are thought to be 
active as radio pulsars for hundreds to thousands of
times longer than strong-field pulsars.  Taking active lifetimes into
account, it appears that there may be comparable numbers of young
pulsars and MSPs in the Galaxy though the birth rate of MSPs is $\sim
10^4$ times smaller.

A second significant difference is that more than 2/3 of MSPs are in
binary systems, while young, strong-field pulsars are largely solitary
objects, a fact which has both theoretical and observational
implications.  Clearly, the evolutionary pathways that gives rise to
MSPs are integrally related to the interaction of the binary stars
(\cite{arcs82}, \cite{rs83}; for a general review see \cite{bh91}).
Moreover, searches for MSPs must confront the additional selection
effects that mitigate against the detection of accelerated pulsars.
As is shown below, detection of the fastest spinning pulsars is
inhibited by any effects that smear out the pulse, such as dispersive
propagation in the interstellar medium.  Orbital motion, uncompensated
for in the surveys we analyze here, also smears out pulses according
to the change in velocity over the duration of the observation and is
therefore most important for short period pulsars in compact binaries
(e.g. \cite{jk91}).

In this paper, we analyze the spatial distribution of MSPs. Our
purpose is to derive the best estimates for MSP population parameters
through careful consideration of survey sensitivities as a function of
pulse period, dispersion measure and other relevant factors. Our
census essentially measures the local number density of MSPs, the fall
off in number density above the Galactic plane and establishes upper
limits on a diffuse, halo-like component of the MSP population.  We
analyze the implications of the spatial distribution for the
kinematics of the MSP population, inferring the diffusive and
impulsive velocity increments suffered.  We compare predictions of the
distribution of proper motions to extant observations, compare the
spatial distribution of LMXBs and MSPs, and describe the importance of
our determination of the MSP kick velocity for binary evolutionary
scenarios. We provide detailed analysis of the influence of selection
effects on the discovery of short period pulsars and pulsars in
binaries.

Our results have immediate relevance in a number of respects.  Recent
MSP surveys have been conducted on the premise that the MSPs are
essentially isotropically distributed around the Sun, at least to the
depths that surveys probe. Our results establish the scale height and
show that Arecibo type surveys see beyond it.

The third way that MSPs differ from slow-spin pulsars 
is in their peculiar space motions. This is one of the main conclusions
of the present paper.
Kinematic evidence (e.g. Dewey, Cordes \& Wolszczan 1988; Cordes \etal
1990;\cite{w94};  Nice \& Taylor 1995; Nicastro \& Johnston 1995)
suggests that MSPs are low velocity objects, with typical transverse
speeds $\simless 100-200$ \kms. Such velocities are much less than
young pulsars, which have an average speed $\sim 500$ \kms (Lyne \&
Lorimer 1994; Cordes \& Chernoff 1996). Our results allow us to put
more stringent, albeit statistical, limits on the MSP velocities than has
hitherto been achieved. 

The use of the spatial distribution of MSPs as an indirect means for
determining their peculiar velocities is more robust than an analysis
of proper motion data.  The primary reason is that the orbits of MSPs
are perturbed significantly from circular motion around the galactic
center so that, given their ages ($> 0.1$ Gyr), corrections for
differential galactic rotation cannot be made.  Arnaud \& Rothenflug
(1981) applied a similar spatial analysis to young, high-field pulsars
as a means for determining their velocities.  (The methodology is
correct but their assumption that high-field pulsars form a steady,
relaxed population is not.  Today we know that $\sim 25$-30\% escape
the Galaxy and that many radio pulsars shut off before traveling to
the limiting distance for detection.)

Our approach differs in several ways from those taken by other
authors.  First, we use a likelihood analysis to provide the best
estimates of the MSP population parameters, to account accurately for
survey selection effects and distance errors,
and to express clearly our physical
assumptions. Second, we restrict our analysis to pulsars with spin
periods $< 20$ ms.  We do so because it appears that these neutron
stars (NS) are distinct from other pulsars that may have undergone
accretion-driven spinup but were left with longer periods (\cite{bh91}). We
also consider them to be distinct from the higher mass NS-NS binaries,
which have pulsars with longer pulse periods (B1913+16, $P=59$ ms
[Taylor \& Weisberg 1989]; B1534+12, $P=38$ ms [Wolszczan 1991]).
Third, our approach includes the effects of interstellar
scintillations, which modulate the pulsar flux density and influence
the rate of detection in surveys. 

The paper is organized as follows.  In \S\ref{sec:surveys} we discuss
sensitivities of pulsar surveys and derive quantities that are needed
in our analysis.  A preliminary attack on the problem is given in
\S\ref{sec:vvmax} where we present a `$V/V_{max}$' analysis, analogous
to that used on quasars and gamma-ray burst sources, in order to
illustrate the uncertainties involved in pulsar surveys.  We derive
the survey likelihood function in \S\ref{sec:like} and apply it in
\S\ref{sec:disk} to a disk-only distribution of MSPs.  In
\S\ref{sec:vfit} we apply the analysis to a disk model based on
numerical integration of NS orbits in the galactic potential.  We
consider a combined disk and diffuse halo-like model in
\S\ref{sec:disk+halo}.
Sections \ref{sec:vel}-\ref{sec:orbits} present the implications of
our results for the MSP birthrate, the origins of MSPs and their
velocities, and for the optimization of MSP surveys.  In
\S\ref{sec:discussion} we summarize the paper.

\section{PULSAR SURVEYS}
\label{sec:surveys}
\subsection{Minimum Detectable Flux $S_{min}$}

A pulsar survey has a minimum detectable flux density $S_{min}$ that 
depends on radiometer noise,  the pulse shape, and details of the Fourier
analysis used to find pulsars.   
In Appendix \ref{app:smin} we derive $S_{min}$ for surveys, 
which includes pulse broadening effects 
(from  interstellar dispersion  and scattering and from detector time constants)
and the effects of flux variations
from interstellar scintillations (Appendix \ref{app:iss}), 
which are strong for the MSP surveys.

\begin{figure}
\plotfiddle{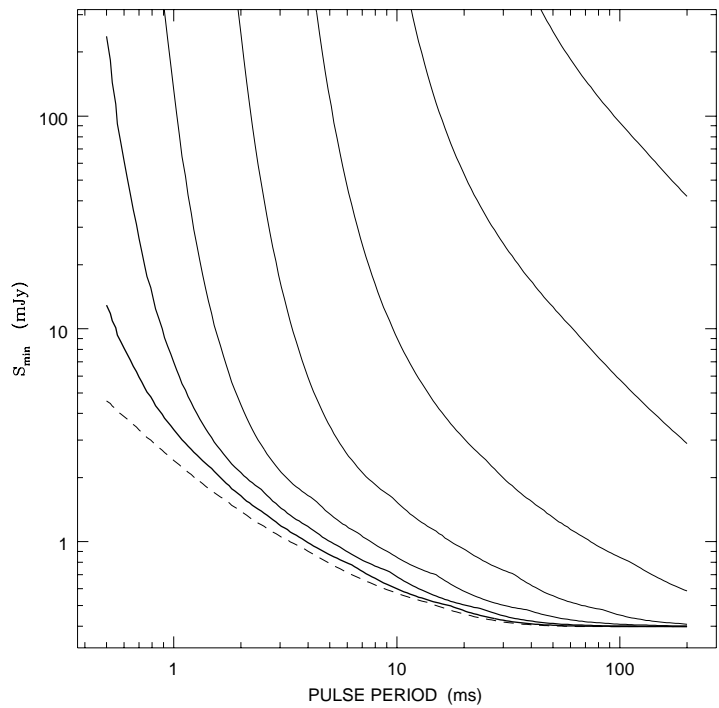}{5.0truein}{0.}{50}{50}{-144}{-72}
\caption{Plot of minimum detectable flux density vs. pulse period for a typical
survey at Arecibo.    The different curves are for dispersion measures
DM $=0$ (dashed line) and $DM= 10\times 2^n$, with $n = 0,1, \ldots$
(solid lines, left to right) and apply to observations
toward the galactic poles and at zero zenith angle.
}
\label{fig:smin}
\end{figure}

For a given direction, the minimum detectable flux density is a function of 
pulse width and period and radiometer-noise level.
The minimum detectable flux density is a function of
direction (galactic coordinates $\ell, b$), dispersion measure
[${\rm DM} = \int_0^D dx\, n_e(x)$, where $n_e$ is the free-electron density and
$D$ is the pulsar distance],   
radio frequency ($\nu$), bandwidth ($\Delta\nu$) and the number of channels
($N_{ch}$), as well as system temperature (T$_{sys})$, 
telescope gain ($G$), the intrinsic pulse duty cycle, and additional
survey dependent factors:  
\begin{equation}
S_{min} = S_{min}(\ell, b, P, DM, \nu, \Delta\nu, N_{ch}, T_{sys}, G, \ldots).
\label{eq:smin}
\end{equation}
$S_{min}$ depends on additional, unspecified parameters, especially those 
that describe orbital motion.  In most of this paper we ignore
such motion; however in \S\ref{sec:orbits} we discuss survey biases
against short period binaries and their possible effects on our conclusions.   
In Figure \ref{fig:smin} we show $S_{min}$ plotted against period for several
values of DM.
These curves apply for drift-scan surveys
made with the Arecibo telescope at 0.43 GHz toward the galactic pole and at 
zero zenith angle. 
We have used a numerical version of a 408 MHz survey (\cite{hssw82}) to
calculate the background sky temperature.
We assume a Gaussian pulse shape with 3\% intrinsic duty cycle. 
While this duty cycle is shorter than those of some MSPs,  
the duty cycle and, hence, $S_{min}$,
is dominated by extrinsic pulse broadening  from dispersion, scattering
and instrumentation.
Pulse broadening from scattering is taken into account  
by estimating it from large scale galactic models for the electron density,
as described by Cordes \etal (1991) and Taylor \& Cordes (1993; hereafter TC).

The maximum distance to which a particular pulsar is
detectable, $D_{max}$, is given by
\begin{equation}
D_{max} = \left [ \frac{L_p}{S_{min}} \right ]^{1/2}. 
\label{eq:dmax}
\end{equation}
We use the `pseudo-luminosity', $L_p \equiv S D^2$, that is
often adopted in population studies of pulsars.  Though it is 
preferable to use a physical luminosity (i.e. expressed in units
of erg s$^{-1}$) in analyzing pulsar
statistics (Chernoff \& Cordes 1996a),
estimation of physical luminosities for MSPs is not yet possible because
we do not understand radio beaming in MSPs to the same extent
that we do for young, strong-field objects 
(Backer 1976; Rankin 1983; Lyne \& Manchester 1988; Rankin 1993).
Note that $S_{min}$ depends
implicitly on the dispersion measure in a given direction which, in turn, 
depends on $D_{max}$.  
For this reason, $D_{max}$ must be found iteratively.    

\begin{figure}
\plotfiddle{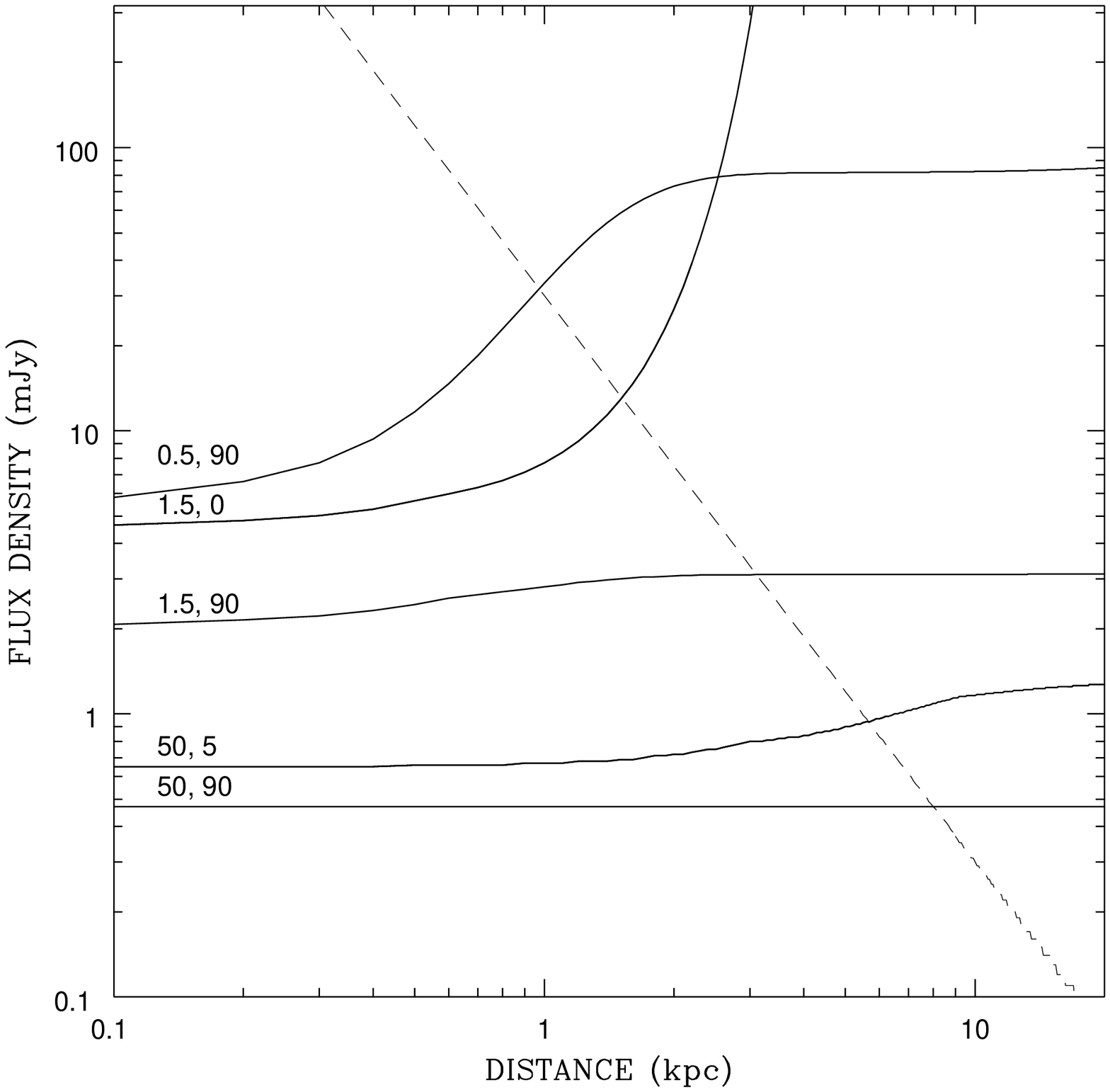}{5.0truein} {0.} {50} {50} {-144} {-72}
\caption{
(Solid lines:) Minimum detectable flux density for an Arecibo survey
for the labelled pulse periods (ms) and galactic latitudes (deg).
The plotted values apply for sources viewed at the telescope beam center.
At the beam half-power point, the values are doubled.
(Dashed line:) The inverse-square law variation of flux density for
a source with 30 mJy when at a 1-kpc distance.
The distances where the solid lines cross the dashed line are
values of $D_{max}$, the maximum distance at which the particular object
could be detected.  At the beam half-power point, the maximum distances
are smaller.
}
\label{fig:smin_vs_d}
\end{figure}

Figure \ref{fig:smin_vs_d}
shows (as solid lines) $\smin$ plotted against distance for several 
values of pulse period and for several directions.  
These curves illustrate the strong dependence of 
$\smin$ on direction  and period.
The dashed line indicates
the inverse square-law variation of flux density for  a source of 30 mJy 
at a distance of 1 kpc. 
The intersection points of the dashed and solid curves 
determine the maximum distances
$D_{max}$ that a pulsar can be seen for the different cases. 

From $D_{max}$, the total volume to which the 
survey is sensitive in  a given beam area  is
\begin{equation}
V_{max} = \frac{1}{3} \Omega_b D_{max}^3.
\label{eq:vmax}
\end{equation} 
Because of the strong period and luminosity dependence of $D_{max}$
and, hence, $V_{max}$, the latter quantity may be used to
determine the period and luminosity distributions of MSPs.
 
\subsection{MSP Surveys: Properties, Volumes and Distances Sampled} \label{sec:app}
 
We have applied the likelihood analysis to 8 pulsar surveys
that have been reported in the literature.   Six use
the Arecibo telescope, yielding 11 MSPs, the seventh is the
Parkes southern-hemisphere survey that yielded 10 MSPs,
as reported by Manchester (1994) when the survey was about
75\% complete.
\footnote{The final tally of the  Parkes survey  is 17 MSPs
with all data analyzed, of which $\sim 95$\% was relatively
free of RFI (M. Bailes, private communication).}
The eighth survey is the Jodrell Bank survey, a portion of which
has been reported in the discovery of the MSP J1012+5307
(Nicastro \etal 1995).  We have not included J0218+4232
(\cite{nbfkl95}) because it was discovered in an aperture synthesis 
survey with selection criteria quite different from the
periodicity searches of the surveys we consider. 
In a future analysis of the distributions
of MSP orbital parameters, we will extend our analysis to synthesis 
surveys. 
 
Table \ref{tab:surveys} lists the surveys we have used.
The columns include the 
(1) survey number (an arbitrary choice
based solely on the order in which we analyzed the surveys);
(2) observatory site for the survey;
(3) survey frequency;
(4) solid angle of interference-free observations;
(5) number of MSPs found in the survey;
(6) system temperature of the telescope, expressed
    in Janskys, for observations at the zenith and toward
    the Galactic poles;
(7) minimum flux density for detection of long-period pulsars
    at the zenith and toward the Galactic poles; and
(8) survey reference number.  

\begin{figure}
\plotfiddle{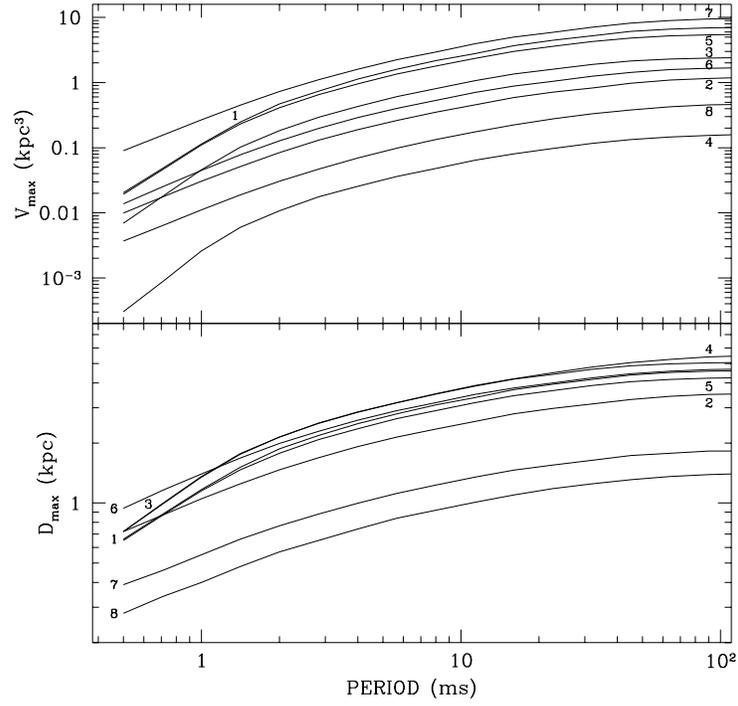}{5.0truein}{0.}{50}{50}{-144}{-72}
\caption{
({\sl Top:})
Volume searched as a function of spin period for a pseudo luminosity
$L_p = $16 mJy kpc$^2$ labelled by survey number as given in
Table \protect\ref{tab:surveys}.
({\sl Bottom:}) Maximum survey distance, $D_{max}$, averaged over
direction for each of the eight surveys and for a luminosity
of 16 mJy kpc$^{-2}$.
}
\label{fig:surveys}
\end{figure}

Table \ref{tab:msplist} lists the MSPs used in our analysis.
Figure \ref{fig:surveys} shows the total volume searched
in each of the 8 programs
as a function of spin period for a fixed luminosity
of 16 mJy kpc$^2$.  The Parkes survey by itself has searched the
largest volume. The Arecibo surveys in aggregate
cover a comparable volume, a few kpc$^3$ for a period of 5 ms.   
For comparison, the lower panel in Figure \ref{fig:surveys},
shows the maximum distance surveyed, $\langle D_{max}\rangle$,
averaged over all directions searched in a given survey. The Arecibo
surveys probe about 3 times more deeply than the Parkes
survey.   

As we demonstrate in this paper, deep, high-latitude surveys at
Arecibo (see references below) sample distances that are well beyond
the scale height of the population, while the southern-hemisphere
Parkes survey, covering a larger area on the sky but being less
sensitive at most periods, is better optimized to finding MSPs.  We
note that the Arecibo search volumes that are devoid of MSPs provide
the most stringent constraints on the scale height of MSPs, whereas
the totality of detected MSPs essentially determines the local MSP
number density.

\section{${\rm V/V_{max}}$ FOR MILLISECOND PULSARS}\label{sec:vvmax}

The main method of analysis in this paper uses a likelihood function to 
determine intrinsic properties of the MSP population after accounting for
survey selection effects embodied in $S_{min}$, as calculated above. 
Here, we motivate our discussion by applying a 
$V/V_{max}$ analysis to MSP surveys.  Consider the line of sight to 
a MSP discovered in a survey.  Subsequent observations yield precise
determinations of P, DM, $\ell$, b.   The flux density is 
$S_d$ at the time of discovery and $S$ as a long time average.   The flux
density is time dependent, owing predominantly  to refractive and 
diffractive interstellar scintillation (DISS; e.g. Rickett 1990;  
Kaspi \& Stinebring 1992, Stinebring \& Condon 1990; 
Cordes, Weisberg \& Boriakoff 1985).   A distance estimate
derives from DM and the TC model for the interstellar electron density.
The model and, in some cases,
auxiliary measurements
(timing parallax,  neutral-hydrogen absorption, and association with
supernova remnants; Frail \& Weisberg 1990) yield
a range of possible distances, $[D_L, D_U]$.   

The volume between us and a given pulsar in a beam of solid angle $\Omega_b$ is
\begin{equation}
V = \frac{1}{3} \Omega_b D^3.
\label{eq:v}
\end{equation} 
The ratio of $V$ to $V_{max}$ from Eq.~\ref{eq:vmax}, 
may then be written as
\begin{equation}
\frac{V}{V_{max}} = \left ( \frac{S_{min}}{S} \right )^{3/2}. 
\label{eq:vvmax}
\end{equation}

Applicaton of Eq.~\ref{eq:vvmax} involves subtleties that depend on 
whether the flux density reported for a given object is influenced 
by the random process associated with DISS.
DISS generally {\it increases} the volume in which a pulsar can be detected
(cf. Appendix \ref{app:iss}).
Saturated DISS that is not quenched by time-bandwidth averaging 
modulates the flux density by a random variable drawn from an exponential 
probability density function (pdf).  
Though the modulation is less than unity more often than not, 
the net effect is to increase the volume by a factor
$\Gamma(5/2) \sim 1.33$. 

\begin{figure}
\plotfiddle{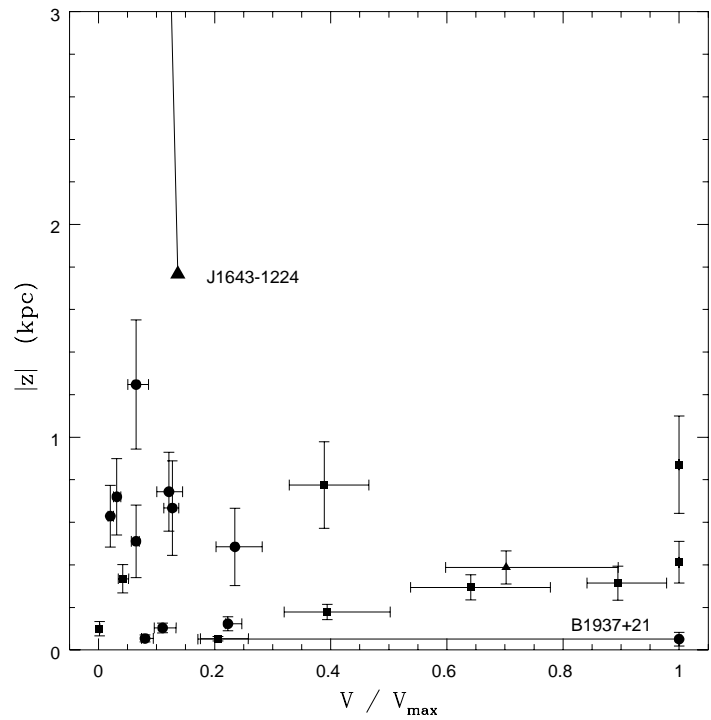}{5.0truein}{0.}{50}{50}{-144}{-72}
\caption{A plot of $\vert z \vert$ vs. $V / V_{max}$ for field
millisecond pulsars.    Filled circles denote pulsars discovered
at Arecibo.  Filled squares indicate MSPs found at Parkes and
Jodrell Bank.  The filled triangle denotes the lower bound on
$z$ and $V/V_{max}$ for J1643-1224, whose distance estimate is
only a lower bound, as discussed in the text.
}
\label{fig:vvmax}
\end{figure}

Figure \ref{fig:vvmax} shows $\vvmax$ for field MSPs plotted against 
$\vert z \vert = D\sin\vert b\vert$. 
Horizontal error bars reflect uncertainties in the measured flux density
and distance errors, the latter also determining the vertical error
bars on $\vert z \vert$.
In most cases, we have used the 400 MHz flux density reported by
Taylor, Manchester \& Lyne (1993), which is usually an average of many
observations and is influenced minimally by DISS.  Flux
calibrations are typically only about 20\% accurate.   Weak pulsars can appear
brighter than average due to DISS at the time of discovery, however, so
the discovery flux $S_d > \smin$ while $S < \smin$.  
For these cases (J0034-0534 and J0711-6830), $\vvmax \to 1$.  
The intrinsically brightest MSP, B1937+21, is detectable
to only $\sim 8$ kpc despite its large luminosity  
$L_p \simgreat 3100$ mJy kpc$^2$  
because, at its low galactic latitude, dispersion and
scattering effects grow rapidly with distance.  Consequently, we argue that
the claimed upper distance limit, $D_U \sim 15.7$ kpc, is a factor of two
too large.  The moderate-latitude pulsar J1643-1224 is attributed only a lower
bound on its distance by the TC model, $D>4.8$ kpc,  because its DM cannot
be accounted for by the model.  We suspect that this pulsar's DM is enhanced
by unmodeled ionized gas along the line of sight  and that the distance is
most likely less than 4.8 kpc.  In the absence of further data, however,
we use the distance lower bound as is.

Apart from the pulsar with a questionable distance estimate (J1643-1224),
Figure \ref{fig:vvmax} shows that MSPs are to be found at only low values of
$\vert z \vert$, suggesting, therefore, that the scale height for MSPs is 
$\sim 0.5$-1 kpc.   To properly estimate the scale height requires careful 
accounting of selection effects in MSP surveys, as we do in \S\ref{sec:like}.  
However, Arecibo surveys at high latitudes search to several kpc for typical 
luminosities.  The absence of high $\vert z \vert$ pulsars is therefore 
especially striking.  The Arecibo MSPs also tend to have small values
of $V/V_{max}$, as would be expected for surveys that search well beyond
the scale height of the population.

\section{LIKELIHOOD ANALYSIS}\label{sec:like}
\subsection{Observables, Assumptions and Statistical Method}

A survey for MSPs typically searches many beam areas for each MSP
discovery.  The spatial distribution of MSPs determines this yield,
along with the survey sensitivity as a function of period, the period
distribution, and the luminosity function.  Here we derive the likelihood
function for a survey, taking these factors into account.
We take as observables the directions of all beam areas searched, the
survey sensitivities in these directions, and the parameters that describe
individual pulsars, including direction, period, flux density, and 
dispersion measure.   We also use distance estimates based primarily on
the electron density model of TC.  Such distances are imprecise and our
method takes into account the large uncertainties in distance that translate
into large uncertainties in implied luminosity. 

Consider a telescope beam with solid angle $\Omega_b$.    The mean number
of MSPs expected in the beam per unit period, luminosity and distance is
\begin{equation}
\frac{\partial^3 \npave}{\partial P\partial L_p \partial D}
   = \Omega_b D^2 n_p(D,\ell,b) f_{L_p}(L_p) f_P(P), 
\label{eq:d3}
\end{equation}
where the number density of MSPs, $n_p(D,\ell,b)$, is an arbitrary
function of position. We have assumed that the joint probability
distribution of period $P$, luminosity $L_p$ and position is {\it
factorable}.  The physical assumption is that the distribution in space
is independent of the distribution of intrinsic pulsar properties ($P$
and $L_p$) and, furthermore, that $P$ and $L_p$ are uncorrelated. We
write the period pdf, $f_P(P)$, and luminosity pdf, $f_{L_p}(L_p)$,
each with unit normalization. With the small number of MSPs currently
available there is scant evidence that the factorization is or is not
appropriate; however, in the future, our method can be applied easily
to more complicated joint distributions if warranted.  In particular,
we defer to another paper exploration of the joint statistics of $L_p$
and $P$. Below, we specialize to disk and disk + diffuse models and we
take into account the variation of the telescope gain across the its
beam.

To calculate the mean number of MSPs expected per beam, we integrate 
Eq.~\ref{eq:d3} to obtain
\begin{equation}
\npave = \Omega_b \int dP f_P(P) \int dL_p f_{L_p}(L_p)
  \int_0^{D_{max}} dD D^2 n_p(D,\ell,b).
\label{eq:npave}
\end{equation}
The volume searched per beam, averaged over $P$ and $L_p$,  is
\begin{equation}
\delta V_{s} = 
    {\Omega_b \over 3} \int dP f_P(P) \int dL_p f_{L_p}(L_p) D_{max}^3.
\label{eq:volsea}
\end{equation}
Survey sensitivities are implicit in $D_{max}$, as discussed in 
\S\ref{sec:surveys}.
For detections, we take into account the
constraints that exist, due to post-discovery observations, on
period, flux density, and distance:
$P\pm\Delta P/2$, $S\pm\Delta S / 2$, and $D\in [D_L, D_U]$. 
Integrating over the subvolume bounded by these constraints, the mean
number of MSPs is
\begin{eqnarray}
\npave_{D} 
       &=& \Omega_b \int_{P\pm\Delta P/2} dP f_P(P) 
          \int_{D_L}^{{\rm min}[D_U,D_{max}]} dD D^4 n_p(D,\ell, b)
          \int_{S\pm\Delta S/2} dS^{\prime} f_{L_p}(S^{\prime}D^2) \nonumber \\
\\
\label{eq:npd}
       &=& \Omega_b  \int_{P\pm\Delta P/2} dP f_P(P)
          \int_{S_1 D_L^2}^{S_2 D_U^2} dL_p f_{L_p}(L_p)
          \int_{{\rm max}[D_L, (L_p/S_2)^{1/2}]}^{{\rm min}[D_U, D_{max}, 
                (L_p/S_1)^{1/2}]} 
               dD D^2 n_p(D,\ell, b), \nonumber
\end{eqnarray}
where $S_{1,2} \equiv S \mp \Delta S/2$.  For each known pulsar,
$D_{max}$ is the maximum distance for the survey that could have
detected each pulsar.  Several MSPs in our analysis were first
discovered in other surveys but were subsequently detected
(rediscovered) in one of the eight surveys and we analyze them
accordingly.  For a given beam, the Poisson probabilities for
detecting zero or one MSP are
\begin{eqnarray}
P_0 &=& e^{-\npave} \cr
P_1 &=& \npave e^{-\npave}. 
\label{eq:poisson}
\end{eqnarray}
We construct the survey likelihood function as the  product
of nondetection (ND) and detection (D) factors:
\begin{equation}
{\cal L} = {\cal L}_{ND} {\cal L}_{D},
\label{eq:like}
\end{equation}
where, for $N_b$ total beams searched, $M_p$ MSPs found, and assuming
$\npave \ll 1$,
\begin{eqnarray}
{\cal L}_{ND} &=& \exp\left( - \sum_{j=1}^{N_b} \npave_{j} \right )
\label{eq:like-nd}
\\
{\cal L}_{D} &=& \prod_{k=1}^{M_p} \npave_{D_k}.
\label{eq:likeparts}
\end{eqnarray}
The log likelihood is
\begin{equation}
\Lambda \equiv \ell n\, {\cal L} =  
     - \sum_{j=1}^{N_b} \npave_{j} + \sum_{k=1}^{M_p} \ell n \,\npave_{D_k}.
\label{eq:loglike}
\end{equation}

The likelihood function may be simplified if we factor the pulsar
number density into a constant $n_0$ times a shape factor:
\begin{equation}
n_p(D,\ell,b) = n_0 h(D, \ell, b),
\label{eq:density}
\end{equation}
where $h(D, \ell, b)$ is dimensionless and has a maximum of unity.
Substituting,  the likelihood function becomes 
\begin{equation}
\Lambda(\thetavec, n_0) = M_p \ell n\, n_0 -n_0 V_d +
       \sum_{k=1}^{M_p} \ell n \, \delta V_p,
\label{eq:loglike2}
\end{equation}
where the vector $\thetavec$ denotes the set of parameters other
than $n_0$.
We define the {\it survey detection volume} as the sum over beams,
\begin{equation}
V_d =  \sum_{j=1}^{N_b} \delta {V_d}_j,
\label{eq:volume}
\end{equation}
where (dropping beam labels)
\begin{equation} 
\delta V_d = \frac{\partial\npave} {\partial n_0}.
\label{eq:dvolume}
\end{equation}
and the constrained subvolume per discovered MSP is
\begin{equation}
\delta V_p =
      \frac{\partial\npave_{D}} {\partial n_0}.
\label{eq:dvp}
\end{equation}
The survey detection volume $V_d$
is the volume searched weighted by the dimensionless MSP space density, $h$.
The expected number of MSPs in a survey is simply $n_0 V_d$.
Eq.~\ref{eq:loglike2} applies to a single-component density model,
such as the disk distribution we consider in the next two sections. 
Multiple components require the alternative treatment
of \S\ref{sec:disk+halo}.

Maximizing ${\cal L}$ with respect to
$n_0$, we obtain the best fit number density
(for a specific set of parameters, $\thetavec$)
\begin{equation}
\n0hat = \frac {M_p} {V_d}.
\label{eq:n0hat}
\end{equation}
Substituting,  the log likelihood becomes
\begin{equation}
\Lambda(\thetavec,\n0hat) = M_p \ell n \, \n0hat - M_p +
     \sum_{k=1}^{M_p} \ell n \, \delta V_p.
\label{eq:loglike3}
\end{equation}
For $n_0\ne \n0hat$, the variation in the log likelihood is
\begin{equation}
\Lambda(\thetavec, n_0) - \Lambda(\thetavec, \n0hat) = 
   M_p \left [ \ell n\left(\frac{n_0}{\n0hat}\right)
   - \left ( \frac{n_0 - \n0hat}{\n0hat} \right ) \right ]
   \approx -\frac{M_p}{2} \left ( \frac{n_0 - \n0hat}{\n0hat} \right )^2,
\label{eq:loglike4}
\end{equation}
where the approximate, quadratic form holds for 
$\vert n_0 - \n0hat \vert/\n0hat \ll1$. 
 
We want to know the marginal distribution
of each parameter.  For a given parameter
$\theta_j\in\thetavec$, the marginal pdf is
the normalized integral over all other parameters
\begin{equation}
f_{\theta_j}(\theta_j) =
   \frac{
       \int_{exc. \theta_j} d\thetavec \int dn_0 {\cal L}(\thetavec, n_0)
        }
        {
       \int d\thetavec \int dn_0 {\cal L}(\thetavec, n_0)
        }
    \approx
  \frac{
       \int_{exc. \theta_j} d\thetavec  {\cal L}(\thetavec, \n0hat) \n0hat
        }
        {
       \int d\thetavec  {\cal L}(\thetavec, \n0hat) \n0hat
        },
\label{eq:thetamarg}
\end{equation}
where the integral subscript `exc. $\theta_j$' means that all parameters
except the j$^{th}$ one are integrated over.
The approximate form in Eq.~\ref{eq:thetamarg} 
assumes a sharp peak about $\n0hat$ and becomes an increasingly
good approximation as $M_p$ grows. 
The marginal pdf for $n_0$ is
\begin{equation}
f_{n_0}(n_0) =
   \frac{
       \int d\thetavec {\cal L}(\thetavec, n_0)
        }
        {
       \int d\thetavec \int dn_0 {\cal L}(\thetavec, n_0)
        }
  \approx
 \left( \frac{M_p}{2\pi}\right)^{1/2}
   \frac{
       \int d\thetavec {\cal L}(\thetavec, \n0hat)
       e^{\displaystyle {-\frac{M_p}{2}\left(\frac{n_0-\n0hat}{\n0hat}\right)^2}}
        }
        {
       \int d\thetavec {\cal L}(\thetavec, \n0hat) \n0hat
        }.
\label{eq:n0marg}
\end{equation}
 
For disk models the areal, or column,  
density of MSPs is less model dependent  
than the number density and scale height separately.  The 
column density is $N_0 \equiv \eta n_0\sigma_z$, where 
$\eta$ is a dimensionless factor of order unity and $\sigma_z$
is a scale-height parameter.  The pdf of 
$N_0$ is calculated by marginalizing ${\cal L}$ 
over all parameters except $n_0$ and $\sigma_z$.  The resultant
joint pdf $f_{n_0, \sigma_z}$ is then integrated according to: 
\begin{equation}
f_{N_0}(N_0) = \eta^{-1} \int d\sigma_z 
     \sigma_z^{-1}f_{n_0,\sigma_z}(N_0/\eta\sigma_z, \sigma_z).
\label{eq:column}
\end{equation}
For disk models considered below, $\eta = 2$ for an exponential in $z$
and $\eta = \sqrt{2\pi}$ for a Gaussian in $z$.

\subsection{Telescope Gain}

MSP surveys usually involve drift scans or sustained pointings
toward specific sky positions.  The telescope's
gain toward a given source varies over the analyzed portion of the drift scan
and is a function of the source's position relative to the beam center
(see, e.g., \cite{cnt96}).   We account for gain variations
by replacing the beam solid angle $\Omega_b$ in Eq.~\ref{eq:npave}
with a sum over equal solid-angle terms
\begin{equation}
\Omega_b \to \sum_{m=1}^{n_g}\delta\Omega_b,
\label{eq:beamgain}
\end{equation}
where the telescope gain varies with $m$, $G_m$.  The minimum
detectable flux density $S_{min}$ is therefore  a function of $m$.
For some drift-scan surveys, we take into account that the data are
analyzed in data blocks that overlap by some fraction (usually 50\%).

For drift scans, $G$ varies with time over the data set and the offset
from the beam center in declination is also taken into account.   The sum
in Eq.~\ref{eq:beamgain} becomes a sum over discrete steps in declination.
For pointed (tracking) observations, we use actual pointing directions
and break the beam into equal-solid angle annuli about the beam center, 
which we sum over as in Eq.~\ref{eq:beamgain}.  We find that only a
small number of subbeam elements is needed to account for the shape
of the beam, e.g. $n_g \sim 2$ or 3.  

\subsection{Interstellar Scintillations}\label{sec:iss}

In Appendix \ref{app:iss} we derive the effects of diffractive interstellar
scintillations on flux densities and on the (pseudo) luminosity function. 
To use these results, we replace
$f_{L_p}$ in Eq.~\ref{eq:npave}
with the corresponding `scintillated' luminosity
function, $f_{\lpprime}$, as defined in the Appendix. 
We do so for surveys assuming that specific sky positions are observed only
once. 
However, we use the unscintillated luminosity function
in Eq.~\ref{eq:npd} because flux densities reported for the known pulsars 
are generally long-term averages  of many independent measurements. 

\subsection{Comparison with Other Statistical Methods}\label{sec:other}

Our statistical method differs substantially from other studies of the
MSP population. A common approach to population studies, including
pulsars and gamma-ray bursts, makes use of nonparametric estimators.
The rationale is to try to draw inferences about certain properties of
the population without assuming a specific class of models. In
contrast, our likelihood analysis makes very specific assumptions
about the class of models to be examined; for example, we have assumed
{\it a priori} that all probability distributions are continuous. The
differences between parametric and nonparametric treatments highlights
some of the strengths and weaknesses of our approach.

It has been shown (Loredo and Wasserman 1995) that nonparametric
estimators may be derived from a special maximum likelihood model
solution. Since our parametric treatment is also based on a maximum
likelihood analysis, it is straightforward to study the relationship
between alternative methodologies by making a comparison of the
assumptions made in the two searches.  The special solution leading to
the nonparameteric estimators of interest comes from a search for
a maximum likelihood solution amongst all functions and generalized
distributions (i.e. delta functions) with equal {\it a priori}
weight. This class of functions is so large that the most likely model
is {\it always} one which exactly and precisely describes the observed
data; thus, nonparametric estimators satisfy the rationale for which
they are introduced. This contrasts with the parametric treatment
adopted here for which the class of functions is (by comparison)
extremely small. We liken nonparameteric estimators to models with
large numbers of free parameters.

In deciding what treatment to adopt, it is helpful to appeal to the
Bayesian odds ratio to decide whether adding a new parameter to a
model is justified by the better description of the data it may
entail. Roughly speaking each newly added parameter will improve the
quality of the model's description of the data. The odds ratio allows
a quantitative decision to be made whether to adopt the more complex
model by weighing the improvement in the description against the
additional freedom to fit arbitrary data sets. The situation for the
MSP's is that the population is rather small and we have anticipated
(without any detailed investigation) that the odds ratio will favor
models with relatively small numbers of parameters.  We have therefore
focused in this paper on parameteric methods with small numbers of
parameters.

An additional factor in our choice of parametric methods is that it is
straightforward to include ancillary information about the population
(e.g. continuity of the model), whereas in nonparametric approaches
such constraints are difficult to incorporate. Moreover, we find the
parametric approach naturally allows the inference of population
parameters of significant interest (e.g. cutoffs in the period
distribution). 

The main drawback of the parametric approach is that the results apply
only to the particular set of models that the parameters can describe.
If the real data were much better described by some completely
different unstudied model, one would have no indication of that fact.
In this paper we have considered several plausible models but these
cannot begin to describe all possibilities.

A number of pulsar population studies are based in whole or in part on
such estimators (Vivekenand \& Narayan 1981, hereafter VN; \cite{pb81}
and \cite{n87}). To be a bit more descriptive, in the VN method a scale
factor is calculated for each object detected in a survey. The factor
represents how many pulsars with the same period P and luminosity
$L_p$ exist in the Galaxy given the fraction of the Galaxy searched.
In our notation, the scale factor is
\begin{equation}
{\cal S}(P,L_p) = \frac
    {\displaystyle \sum_{\rm full\,\,sky} {\Omega_b}_j
                 \int_0^{\infty}  dD\, D^2 n_p(D,\ell,b) }
    {\displaystyle \sum_{j=1}^{N_b} {\Omega_b}_j 
                 \int_0^{D_{max}} dD\, D^2 n_p(D,\ell,b) },  
\label{eq:scalefactor}
\end{equation}
where $D_{max}$, as before, depends on many survey and pulsar parameters,
including $P$ and $L_p$. 
The number of pulsars in the Galaxy is then calculated through a
sum over detected pulsars as
\begin{equation}
N_{gal} \approx \sum_{i=1}^{N_{msp}} {\cal S}(P_i, {L_p}_i) .
\label{eq:ngal}
\end{equation}
The resultant total number of pulsars is a {\it mean value} similar in
nature to the mean value of the number density, $\hat n_0$, that we
have calculated. One drawback is that the VN method estimates the
number of pulsars in the Galaxy {\it exactly} like those actually detected.
In other words, it explicitly includes contributions to the mean only
at the periods and luminosities of the known pulsars. It is inherently
discrete as compared to our likelihood method based on continuous
distributions. Another drawback is that the method does not directly
allow computation of confidence intervals.  Finally, since the scale
factors are calculated only for the detected pulsars, there is no
means for estimating the cutoffs of the distributions of $P$ and
$L_p$.  Below we compare our results on MSPs to those of 
Lorimer \etal (1995) and Bailes \& Lorimer (1995; hereafter BL) 
with these issues in mind.

\section{DISK MODEL FOR MILLISECOND PULSARS}\label{sec:disk}
\subsection{Method}

The simplest spatial model is a disk with constant scale height
$\sigma_z$, so that the density is a function of z only.     
We let $n_p(z) = n_d h_p(z)$ with $h_p(0) = 1$, where
$n_d$ is the midplane density that corresponds to $n_0$ in
\S\ref{sec:like}.

The parameters to be solved for describe the period, luminosity and
$z$ distributions, $f_P(P)$, $f_{L_p}(L_p)$ and $n_p(z)$.
We have considered three models for $h_p(z)$: 
(1) a Gaussian function in $z$ with an rms value of $z$ given by $\sigma_z$;  
(2) an exponential model with 1/e scale height $\sigma_z$; and 
(3) a numerically derived distribution of NS orbits, neither Gaussian nor
exponential in form, discussed in \S\ref{sec:vfit}.   
For the luminosity and period pdfs, we adopt power-law functions, i.e. 
$f_{L_p} \propto L_p^{-\alpha_{L_p}}$ and
$f_P \propto P^{-\alpha_P}$,
with respective lower and upper cutoffs, ${L_p}_1, {L_p}_2$  and $P_1$, $P_2$.

The greatest computational effort goes into calculation of 
${\cal L}_{ND}$ (Eq.~\ref{eq:like-nd}).   
We computed it efficiently by summing  the
D integral in Eq.~\ref{eq:npave} over the survey beam areas for 
a grid of $\sigma_z$, P, and $L_p$; as stated before, we use
the scintillation-modified luminosity function in this computation. 
Next we form the likelihood for  different model parameters 
$\alpha_{L_p}, \alpha_P, {L_p}_1, {L_p}_2, P_1, P_2$ 
by calculating integrals over P and ${L_p}$ with
weights $f_P$ and $f_{L_p}$ (cf. Eq.~\ref{eq:npave}).

We maximized $\Lambda$ by varying the parameters (or subsets of the
parameters) over a grid. We kept the upper cutoffs on the period and
luminosity distributions fixed at $P_2 = 20$ ms and ${L_p}_2 = 16,000$
mJy kpc$^2$. The period cutoff corresponds to the selection used to
define the sample. Since the number of objects decreases rapidly as
$P$ increases, the upper cutoff plays little role in any of the
results below. The luminosity cutoff corresponds to the maximum
possible luminosity in the observed sample.  We also tested the
effects of varying ${L_p}_2$ and found that results are not sensitive
to this parameter.  Exclusion of B1937+21, the most luminous MSP,
allows a much smaller value for ${L_p}_2$ to describe the remaining 21
pulsars in the sample; but none of the other results below are
substantially altered.  

\subsection{Results}

\begin{figure}
\plotfiddle{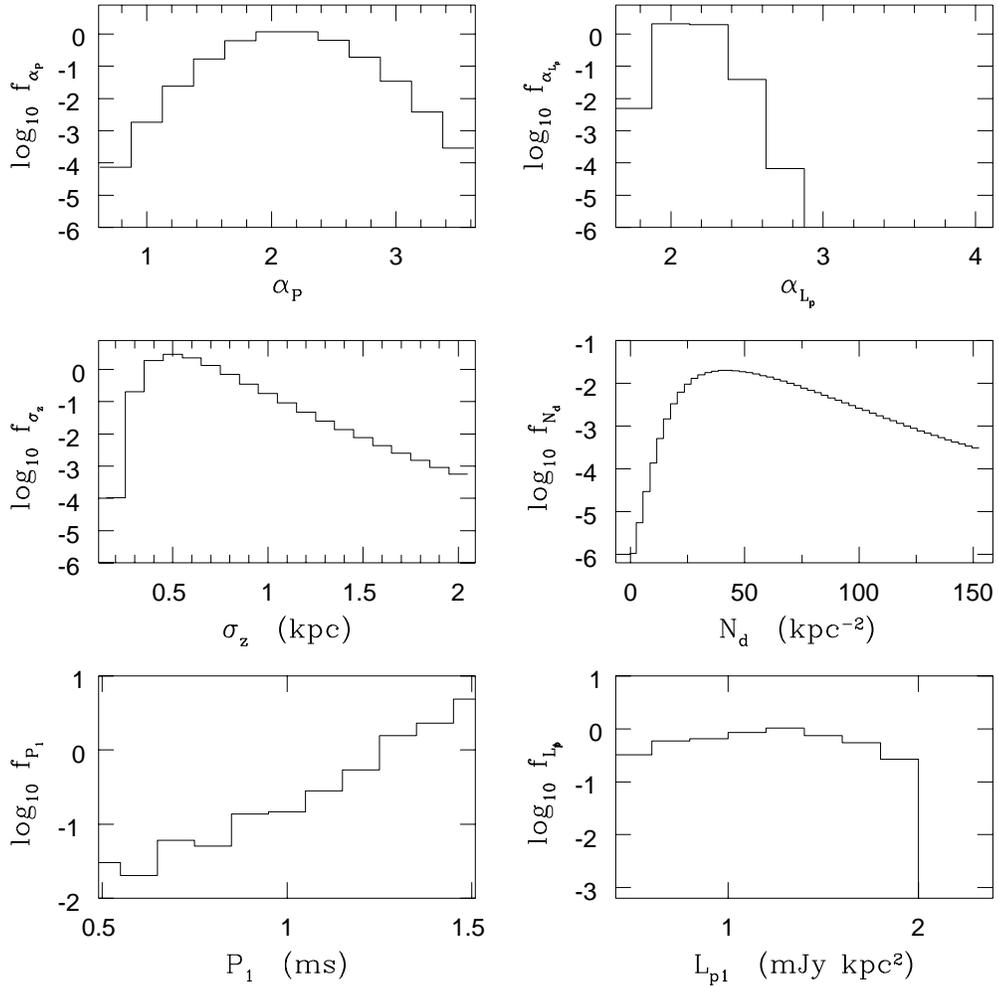}{6.0truein}{0.}{70}{70}{-200}{-72}
\caption{
Marginal pdfs for the 6 parameters of the exponential disk model for
the millisecond population.  $N_d$ is the column density (kpc$^{-2}$)
of MSPs, which we show instead of the number density, $n_d$.
}
\label{fig:marg}
\end{figure}

The five parameters ($P_1$, ${L_p}_1$, $\alpha_{L_p}, \alpha_P$, and
$\sigma_z$) were varied over a grid to find the maximum $\Lambda$.
We formed marginal pdfs according to 
Eqs.  \ref{eq:thetamarg} and \ref{eq:n0marg}.
Results are summarized in Table \ref{tab:bestfit}.  Figure
\ref{fig:marg} shows the marginal pdfs for each of the six parameters
(the above-mentioned five and the number density, $n_d$).  Using these
pdfs, we calculated the confidence intervals on the parameters that
are given in Table \ref{tab:bestfit}. The maxima are well-defined and
easily located.

\subsubsection{Minimum Period $P_1$ and Period Distribution Slope $\alpha_P$}

Naturally $P_1$ must be less than or equal to the period of the
shortest-period MSP in our sample.  When other parameters are held
fixed, it is straight forward to show that $\Lambda$ must decrease as
$P_1$ is made smaller. The best-fit, minimum period lies only slightly
below that of the most-rapidly-spinning, known pulsar, B1937+21 (1.56
ms).  However, the data allow $P_1 < 1.56$ ms at a reduced level of
confidence.  The results are given in Table
\ref{tab:bestfit} for both the Gaussian and exponential models.
The cutoff is $>1$ ms at 95\% confidence and $>0.65$ ms at 99\% confidence.

The period distribution falls off steeply with period, implying the
existence of many objects at small $P$ ($dN/dP \propto f_P \propto
P^{-2.0\pm0.33}$).  It is well known that physical instabilities will
act on neutron stars with very short rotation periods.  Ignoring the
magnetic field and assuming accretion from an inner edge of a
Keplerian disk, Cook \etal (1994a,b) have shown that $1.4 \msun$
neutron stars can be spun up to critical rotation periods (well under
1 ms) for a variety of equations of state without triggering radial
instability, e.g. exceeding the maximum neutron star mass. (The
results do not assure stability against non-radial modes and the
associated gravitational wave emission.) Our overall fit for the
period distribution suggests the existence of MSPs faster than those
that are currently known (1.56 ms) in view of the fact that the
theoretical stability analyses do not rule out such objects. Of
course, there may be evolutionary reasons that such objects do not
occur and we discuss the significance of the cutoff $P_1$ next.

The specific value of $P_1$ depends, of course, on our assumption of
a power-law distribution for $P$.  We have not explored other mathematical
forms, but reasonable alternatives include a power law that flattens
for periods less than some critical period and cuts off at $P_1$
or a distribution that rises slowly from zero at $P_1$ and peaks
at or near 1.56 ms and then follows a power-law form like that we have
fitted.   It is easy to see that such alternative period distributions
will lead to {\it smaller} $P_1$ than we have derived.  The reason is
that they imply that smaller volume has been surveyed for $P<1.56$ ms,
so the allowed $P_1$ can be smaller.   Therefore, our derived
$P_1$ using the power-law distribution is a maximally allowed value
and suggests, conservatively, that the period range for MSPs
may extend to as small a value as 1 ms (95\% confidence) or
0.65 ms (99\% confidence). 

Harding (1984) analyzed the slope of $f_P$, assuming a steady-state
flux with births balanced by pulsars crossing the Hubble line.  She
showed that if pulsars are born with a powerlaw distribution of $B$
($\propto B^\beta$) and with initial period $P$ approximately $\propto
B$ (accretion spin up models imply $B^{6/7}$) then the resultant $f_P
\propto P^\beta$.  Today it is known that the spin-down times for the
observed MSPs are too long for a steady-state to be attained.
However, with similar assumptions we find the same slope in the period
range $[P_{min}(T_h),P_{max}]$, where $P_{min}(T_h)$ is the period
reached after a Hubble time ($T_h$) by the minimum initial period
object (minimum magnetic field) and $P_{max}$ is the longest period at
birth.  (Different slopes are found in other period subintervals.
Additional discussion will be found in Chernoff \& Cordes 1996a).
Thus, one possible interpretation of the steep period slope is that
the field distribution $\propto B^{-2.0\pm0.33}$.  However, it is
difficult to derive robust constraints on the field distribution
without knowing both the time dependence of magnetic fields during the
spinup process and the spindown law for MSPs subsequent to the
spinup phase.  

\subsubsection{Scale Height $\sigma_z$}

The inferred Gaussian scale ($0.65$ kpc) and exponential scale ($0.50$
kpc) are in rough agreement. The values indicate that the MSPs have a
relatively small scale height, comparable to the oldest disk stars.
Though the confidence intervals overlap, the actual shape of the
distribution plays some role in the value of the scale height parameter
and motivates, in part, a more physical analysis based on motion
of objects in the Galactic potential (\S\ref{sec:vfit}).

\subsubsection{Minimum Luminosity $L_{p1}$ and Slope $\alpha_{L_p}$}

The luminosity pdf of our best fit, $dN/dL_p \propto f_{L_p}(L_p)
\propto L_p^{-2\pm0.2}$, is similar to that of long-period pulsars
(e.g. Lyne, Manchester \& Taylor 1985). Total numbers are dominated by
weak sources. The lower cutoff is $L_{p1} = 1.1^{+0.4}_{-0.5}$ mJy 
kpc$^{-2}$ and is largely determined by the absence of nearby sources.
We have shown for long-period pulsars that $f_{L_p}$ is
strongly influenced by geometrical beaming effects, the distribution
of true luminosities, the spin down law and a death line (Chernoff \&
Cordes 1996a). Because all four of these elements may differ
between high-field pulsars and MSPs, we currently regard the 
similarity between the long-period and MSP luminosity 
distributions as fortuitous.

In the past, the disk-determined $f_{L_p}(L_p)$ (slope and cutoff) has
also been used to make inferences about the number of MSPs in globular
clusters. On evolutionary grounds, many properties
of disk and globular cluster MSPs might be expected to differ 
(e.g. distributions of luminosity, spin period, orbital period and velocity).
Since the nearest cluster is too distant to allow direct measurement
of the luminosity function near $L_{p1}$, usage of the disk-determined
form is necessary for many purposes. Fruchter \& Goss (1990) measured
the radio flux from nearby globular clusters and estimated
$\sim 10^3$ MSPs in the Galaxy's globular cluster system.  Our best
fit luminosity distribution, with cutoffs, is consistent with the one
they assumed and does not alter the size of this estimate. Likewise,
estimates by 
Foster \& Tavani (1992) and Johnston, Kulkarni \& Phinney (1992)
of the {\it shape} of the luminosity pdf for MSPs in globular
clusters are also consistent with our best-fit $f_{L_p}$ for disk
MSPs, though both groups were unable to determine the 
lower luminosity cutoff and, hence, the absolute normalization. 
Wijers \& van Paradijs (1991) find far fewer globular cluster
MSPs than do Fruchter \& Goss or Johnston, Kulkarni \& Phinney
even though they adopted a lower luminosity cutoff three times smaller 
than that of Fruchter \& Goss; the difference is probably related
to their assumed dependence of luminosity on
spin period and spin period derivative that was based on 
young, high-field pulsars.    Analysis of globular
cluster MSP populations should probably use a treatment similar 
to this paper's but applied to cluster-only data.  

\subsubsection{Correlations}

\begin{figure}
\plotfiddle{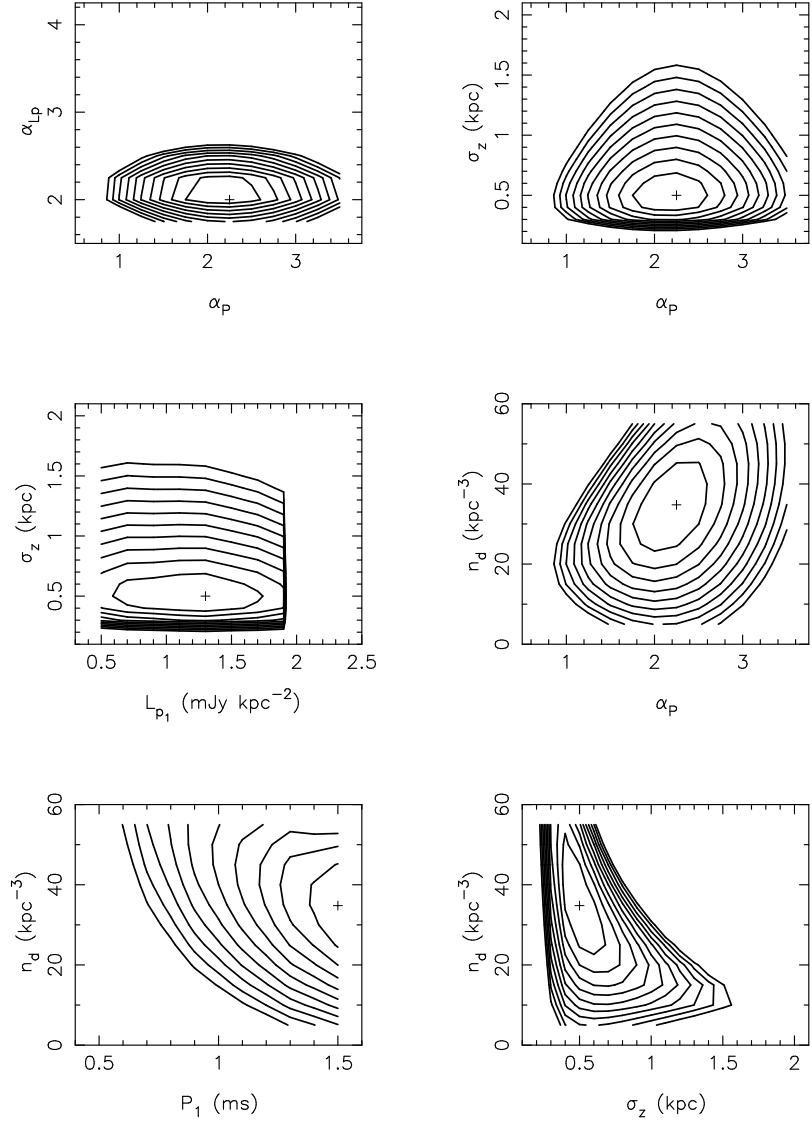}{7.0truein}{0.}{60}{60}{-144}{+00}
\caption{
Selected contour plots of the log likelihood for the exponential model
plotted against pairs of
parameters while holding the other four parameters fixed at their
values that yield the maximum likelihood.  Contour spacings are unity
in natural log units and the first contour is a factor 1/e from the peak.
}
\label{fig:contours}
\end{figure}

Most of the derived parameters are uncorrelated.  However, $\alpha_P$
and $P_1$ are positively correlated as are $\alpha_{L_p}$ and
${L_p}_1$, while $n_d$ is negatively correlated with the lower cutoffs
in period ($P_1$) and luminosity (${L_p}_1$).  Figure
\ref{fig:contours} shows contours of constant likelihood plotted
against pairs of parameters while holding all other parameters fixed
at values that yield the maximum likelihood.

\section{DYNAMICAL MODELS}\label{sec:vfit}
\subsection{Birth Kick Determination}

In \S\ref{sec:disk} we assumed functional forms for the $z$
distribution of MSPs and fit for the associated scale height
parameters.  These parameters describe the present-day MSP
distribution without regard to the orbit about the Galaxy.
We have constructed a
dynamical model that connects ``birth parameters'' to today's spatial
distribution as follows.  We model the birthrate density of MSPs
\begin{equation}
{\dot n}(R,z) = g(R) \exp\left(-{z^2 \over 2 \sigma_{z,b}^2}\right),
\end{equation}
where $R$ is (cylindrical) Galactocentric radius,
$z$ is height about the plane, $\sigma_{z,b}$ is a scale height parameter
and $g(R)$ is a surface density function, taken to be either constant
(``uniform model'') or exponential with scale length 3.5 kpc (``exponential
model'').

The birth velocity is the circular rotation velocity plus a kick
component.  Note our use of ``kick'' includes {\it any} momentum
impulse imparted to the pulsar's progenitor or companion, if in a binary. 
  The angular distribution of the kick is isotropic
and the velocity magnitude has a distribution 
$\propto V^2 e^{-V^2/2\sigma_{V}^2}$. After birth,
the MSP trajectory is determined by integration of the orbit about the
Galaxy in a simplified model of the gravitational potential
(Pacyzynski 1990).  We ignore the role of scattering from
irregularities (e.g. GMC's, spiral density waves, massive black holes)
in the calculated motion. We first discuss the kinematic properties of
the MSP population inferred from the smooth model and next assess the
degree to which our conclusions may be modified by the diffusion of
stellar orbits.

About 4 million orbits were integrated over time spans of $10^9$
years, sufficiently long that the derived vertical distribution was
stationary and well-mixed. For specific birth parameters, the vertical
distribution of MSPs in the vicinity of the Sun (e.g. in an annulus
of Galactocentric radii from $7.5-9.5$ kpc) 
was calculated by appropriately weighting and
combining the results for the individual orbits.

\begin{figure}
\plotfiddle{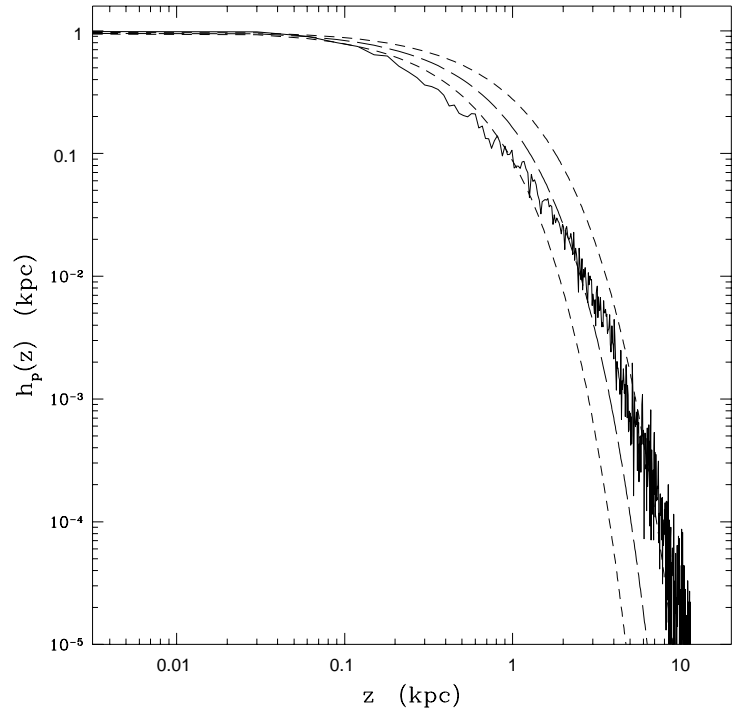}{5.0truein}{0.}{50}{50}{-144}{-72}
\caption{Comparison of the $z$ distributions for the best fit
velocity model (noisy solid line)  and three exponential models
with scale heights of 0.41, 0.55 and 0.78 kpc (dashed lines).
}
\label{fig:zdistributions}
\end{figure}

The statistical analysis described in previous sections was carried
out to determine the birth parameters ($\sigma_V$ and the intrinsic
pulsar population parameters) of the uniform model. The results (Table
\ref{tab:bestfit}) give a peak value $\sigma_V = 52^{+17}_{-11}$
km s$^{-1}$.  The initial scale height is not well-determined by the
data and was held fixed, $\sigma_{z,b} = 0.1$ kpc.  The column density
of MSPs was calculated from Eq.~\ref{eq:column}.
The result is consistent with values obtained using the
Gaussian and exponential spatial models.  Figure
\ref{fig:zdistributions} illustrates the density distribution 
vs. $z$, comparing the range of allowed exponential fits
(\S\ref{sec:disk}) to the most likely dynamical model. The differences
are subtle and suggest that the assumed exponential form should be an
adequate local description for many purposes.

\subsection{Kinematics of Today's Population}

\begin{figure}
\plotfiddle{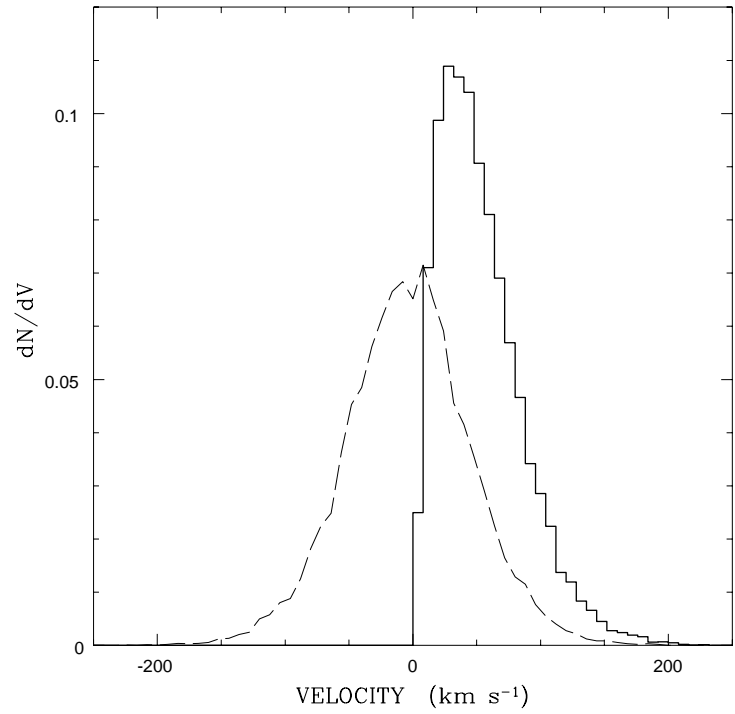}{5.0truein}{0.}{50}{50}{-144}{-72}
\caption{
Histograms of observed transverse speeds (solid)
and line-of-sight (dashed) velocity for the best fit
velocity model of \S\ref{sec:vfit}.
}
\label{fig:pmsim}
\end{figure}

Kinematic properties of the MSP population may be inferred from the
dynamical model. For example, the distribution of parallel and
perpendicular velocities relative to the LSR are easily derived from
the orbital calculations.  Figure \ref{fig:pmsim} shows the
distributions for all simulated objects within 1 kpc of the Sun for
the most likely dynamical model.  The expected transverse motions are
small; approximately 99\% of the MSPs have $\Delta V_{\perp} < 150$ km
s$^{-1}$. As MSP samples increase in number, detailed distributions
like these will provide important additional constraints on modeling.

Next, we consider the velocity ellipsoid of MSPs.
Let $v_R$, $v_t$ and $v_z$ be the components of velocity in the
cylindrical radial direction, in the tangential direction (parallel to
the local circular velocity, e.g. $l=270^\circ$ at the solar position) 
and out of the plane, respectively. In
the well-mixed state, the most interesting non-zero moments are
$<v_R^2>$, $<v_t^2>$, $<v_z^2>$ and $<v_t>$.  Table
\ref{tab:kinematics} lists the first and second moments for a range of
MSP birth models with $\sigma_V = 20$, $40$, $60$, $80$ and $100$
km s$^{-1}$, for two scale heights $\sigma_{z,b} = 0.05$ and $0.15$ kpc, and for
the uniform and exponential surface density distributions.  (All
velocity moments are given in units of $\sigma_V$.) When the kick
velocity is small compared to the rotation velocity and when disk
properties do not vary significantly over the range of radii sampled,
the results of epicyclic theory are directly applicable. For local
objects $<v_t^2>/<v_R^2> = | B/(A - B) |$ has the
observed value $0.45 \pm 0.09$ (for Oort constants $A = 14.5 \pm 1.5$
km s$^{-1}$ kpc$^{-1}$ and $B = -12 \pm 3$ km s$^{-1}$ kpc$^{-1}$ [\cite{bt87}]). The
model-calculated value of $0.45$ at $\sigma_V = 20$ km s$^{-1}$ is in good
agreement with the value inferred from the observed Oort constants.
Here, we will concentrate on the changes that occur as
$\sigma_V$ increases and that are indicative of some of the
differences between the velocity distributions of MSPs and of disk
stars.  A global model is necessary since a local epicyclic treatment
for the MSPs is not well-founded.  For example, with a kick of 60 km s$^{-1}$
particles observed at the local position could come from initial radii
in the approximate range $(0.5-2.6) \times \Rgs$, where $\Rgs$ is the Sun's
distance from the Galactic Center.  Also, kicks of this
size create vertical excursions of $>0.3$ kpc, spoiling a harmonic
approximation to the potential.

Table \ref{tab:kinematics} shows how the basic moments change as
$\sigma_V$ increases. We briefly note the most important conclusions:
(1) A clearly noticeable effect is the occurrence of a non-zero
tangential motion measured with respect to the local circular velocity
(``asymmetric drift'').  For $\sigma_V=60$ km s$^{-1}$ the magnitude is $\sim
13$ km s$^{-1}$ in the uniform model ($\sim 25$ km s$^{-1}$ in the exponential
model), an effect that is potentially detectable in a relatively small
sample of objects with well-determined velocities.  (2) The velocity
ellipsoid (with axial ratios 
$\sqrt{<v_R^2>}$:$\sqrt{<v_t^2>}$:$\sqrt{<v_z^2>}$)
becomes rounder as the magnitude of the kick grows.  (3) The
birth distribution in Galactocentric radius affects
the value of all the non-zero velocity moments including the shape of
the velocity ellipsoid and the magnitude of the asymmetric drift. (4)
The imprint of the birth scale height is essentially absent for
objects with $\sigma_V \simgreat 60$ km s$^{-1}$.

Determinations of asymmetric drift would provide valuable information
on the birth locations of MSPs.  Proper estimation of the effect
will require more field MSPs than are currently known and careful
treatment of distance errors.  We defer a detailed discussion to 
another paper. 

\subsection{Orbital Diffusion}

The model calculations presented above assume a regular background potential.  
Older stars are well known to have larger velocity dispersions, presumably
from interaction with small-scale fluctuations in the gravitational field,
but the actual physical source of the irregular field is not well
understood (\cite{w77}). The oldest stars, K and M giants of age $9
\times 10^9$ yrs, reach total dispersions of 77 km s$^{-1}$; for comparison,
using interpolated values for the uniform model in Table
\ref{tab:kinematics} we estimate that the best fit model for the MSPs
($\sigma_V = 53$ km s$^{-1}$) implies a total dispersion of 84 km
s$^{-1}$.  In fact, the MSPs suffer comparable energy input from kicks
and from diffusion.  The key assumption is that the MSP population
includes members with ages ranging uniformly up to the age of the
Galaxy, so that the average effect of diffusion will be less than it
is for the oldest stars. Using the velocity dispersion data of K and M
giants with ages $(0.3-9) \times 10^9$ (\cite{w77}), averaging uniformly
in time, we infer that the root mean square dispersion is $\sim50$ km
s$^{-1}$.  We suggest that the residual dispersion of $67.5$ km
s$^{-1}$ (i.e.  $\sqrt{ 84^2 - 50^2}$) is due to kick(s) unique to MSP
evolution. This 3D dispersion would then correspond to a 1D kick of
$\sim39$ km s$^{-1}$.

\subsection{Conclusions}

The best fit uniform model implies $\sigma_V = 53$ km s$^{-1}$;
this is an upper limit because gravitational scattering processes 
are ignored in its estimate; 
the scale of the kick is $\sim40$ km s$^{-1}$ assuming MSPs are
long-lived and born at a uniform rate. If MSPs are visible for less than
a Hubble time, the kick size will increase; if most of today's MSPs
were formed early in the Galaxy's life, the kick size will decrease.

\section {DISK + DIFFUSE MODEL}\label{sec:disk+halo}
\subsection{Method}

The MSP distribution may be more complex than a single disk component
with small scale height.  For example, there may exist a population of
MSPs that fill a halo-like region around the disk.

If MSPs are distributed in two components, the log likelihood
becomes
\begin{equation}
\Lambda(\thetavec, n_d, n_h) = -[n_d V_d + n_h V_h] +
 \sum_{k=1}^{M_p} 
     \ell n \, \left [n_d \delta V_p^{(d)} + n_h \delta V_p^{(h)} \right ].
\label{eq:loglikehalo}
\end{equation}
Here, we label disk quantities with `d' 
while `h' denotes diffuse (halo-like) contributions; we suppress 
the dependences of the volumes on other parameters.
Maximizing
$\Lambda$ with respect to $n_d$ and $n_h$, we find that the best-fit
number densities $\hat n_d$ and $\hat n_h$ satisfy
\def\ndhat{{ \hat n_d }}
\def\nhhat{{ \hat n_h }}
\begin{equation}
\ndhat V_d + \nhhat V_h = M_p.
\label{eq:densitysum}
\end{equation}
Also, if we take the $n_h = 0$ case as a fiducial solution,
which is our result in \S\ref{sec:disk} 
for the disk-only model,
the log likelihood for
$n_h \ne 0$ may be expanded as
\begin{equation}
\Lambda(\thetavec, n_d, n_h) = \Lambda(\thetavec, n_d, 0)
    -n_h V_h + \sum_{k=1}^{M_p} \ell n\, 
    \left ( 1 + \frac{ n_h \delta V_p^{(h)}}{n_d \delta V_p^{(d)}} \right ). 
\label{eq:withhalo}
\end{equation} 

\subsection{Diffuse Models}

One might expect a diffuse distribution of MSPs for any of several
reasons.  (1) The probability distribution of birth velocities may
extend to values much larger than typically allowed by the assumed
Gaussian or exponential forms. The high velocity MSPs would oscillate to higher
z distances or escape the Galaxy all together.  (2) MSPs born in
globular clusters may be ejected by dynamical interactions or when a
cluster is tidally dissolved. Such objects would have a spatial
distribution like the parent systems assuming the ejection velocities
were small compared to the rotation velocity.  (3) Spheroid stars may
evolve and produce long-lived MSPs just like disk stars (e.g.  by
accretion-induced spinup). Such objects would have a spatial
distribution like the Pop II spheroid. (4) If the formation of the
Galaxy involved hierarchical merging of smaller objects containing
disk-like structures, their MSPs will be cannibalized.  Such objects
might follow the dark matter halo distribution.

Without further considering the merits of these basic scenarios, we
will adopt several geometrical distributions for the the putative
diffuse population and place upper limits on the number densities.
Consider a density model for MSPs of the form
\begin{equation}
n_h(r) = n_h \left [ 1 + (r/r_h)^2 \right ]^{-s_h/2},
\label{eq:halo_density}
\end{equation}
where $r$ is the radius from the center, $r_h$ is the characteristic
radius, and $s_h$ is the power-law index.  Taking $s_h=0$ gives a
uniform density halo, our reference model (in practice all
distributions are truncated at $50$ kpc).  Taking $s_h=2$ and $r_h =
5$ kpc gives an isothermal distribution with large core.  Taking
$s_h=3.5$ and $r_h = 1$ kpc gives the observed globular cluster
distribution (\cite{t89}).

Using these models, we calculate the diffuse pulsar density by
integrating Eq.~\ref{eq:npave} and we evaluate the halo volume factors
$V_h$ and $\delta V_p^{(h)}$ (Eqs.  ~\ref{eq:volume} and
~\ref{eq:dvp}). We combine these with the analogous disk quantities
and examine a grid in $n_d$ and $n_h$ to find the distribution of
likelihood values.

\subsection{Results}

\begin{figure}
\plotfiddle{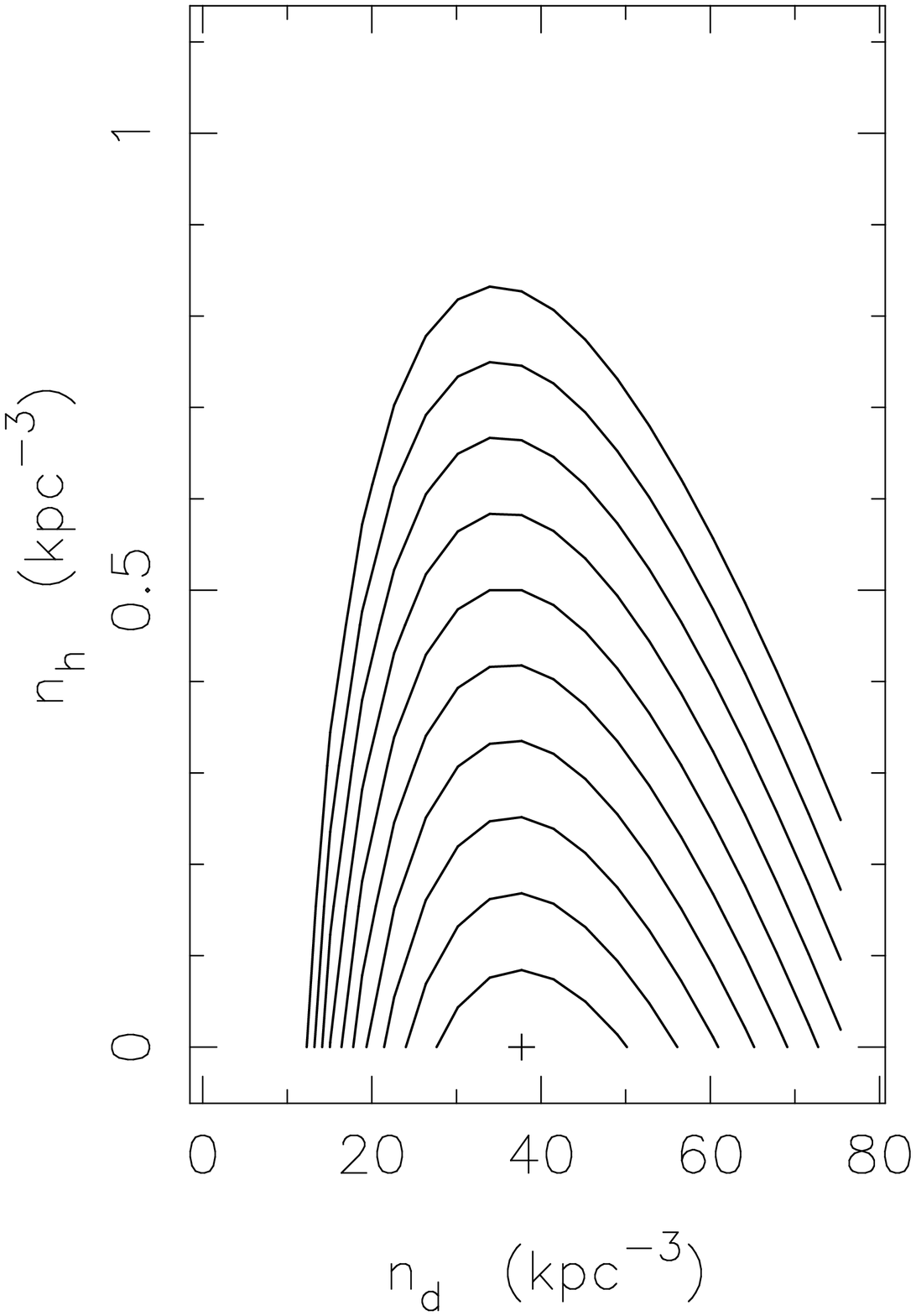}{6.0truein}{0.}{60}{60}{-200}{+00}
\caption{Contours of log likelihood plotted against the densities
of MSPs in disk ($n_d$) and diffuse ($n_h$) components.  The plot is for
a uniform halo that extends well past the solar circle.  Contour spacings
are unity in the natural log.  The plus sign marks the peak
likelihood.
}
\label{fig:halo_contour}
\end{figure}

\begin{figure}
\plotfiddle{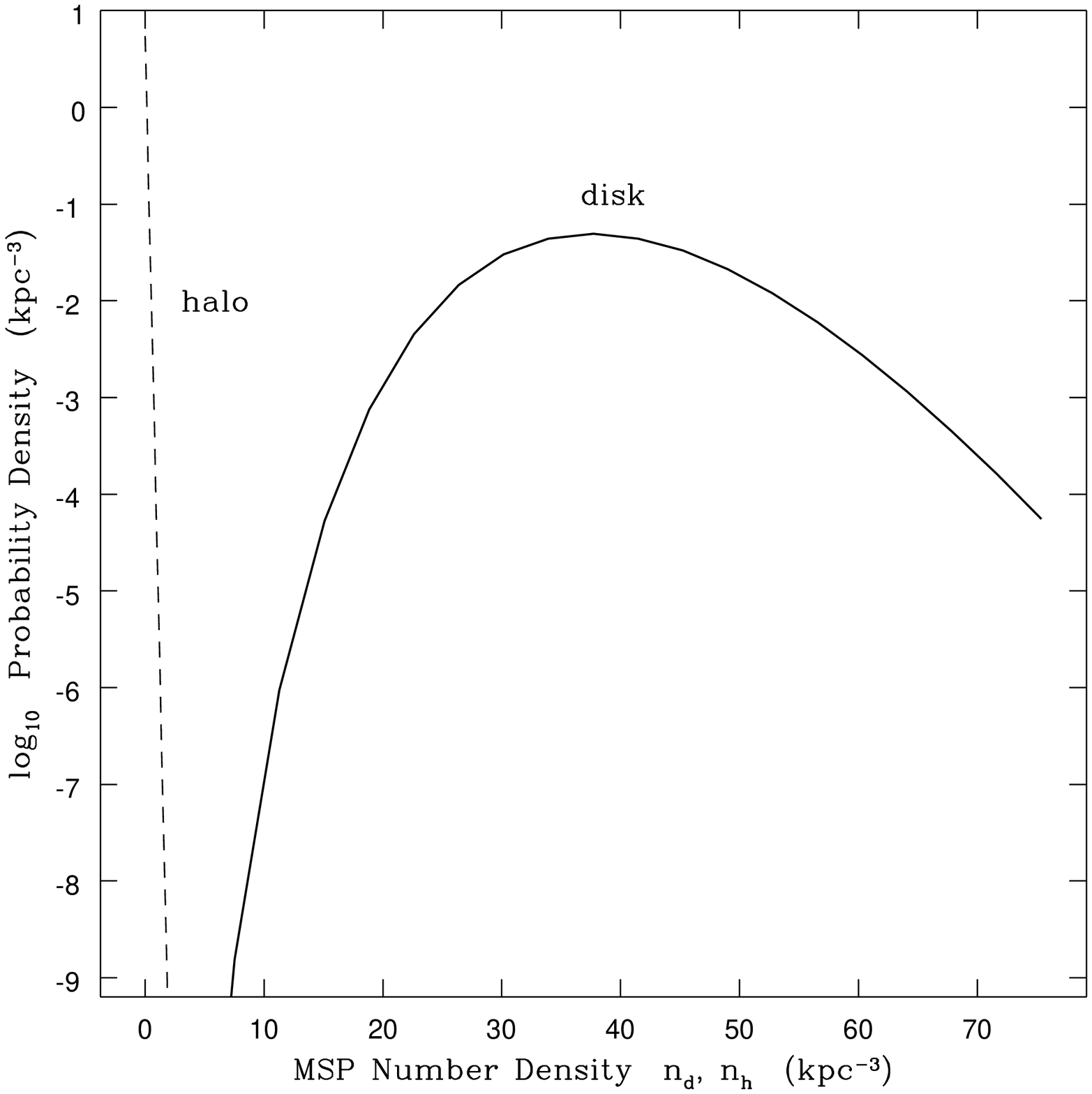}{6.0truein}{0.}{50}{50}{-144}{-72}
\caption{Marginal probability density functions for the disk and
diffuse MSP densities, $n_d$ and $n_h$, respectively.
}
\label{fig:halo-marg}
\end{figure}

For our reference model, we find that the pure disk model is favored
by a huge factor implying an upper bound on the diffuse density from
the fitting is $n_h \simless 0.4$ kpc$^{-3}$ (90\% confidence, cf.
Table \ref{tab:disk_halo}).  Figure \ref{fig:halo_contour} shows
likelihood contours for the disk and diffuse densities, with a maximum
at $n_h = 0$.  The marginalized densities are shown in Figure
\ref{fig:halo-marg} for the uniform density model.

For the other two models, we have expressed the results in terms of
limits on the density parameter $n_h$ (the value at the center of the
Galaxy) and, equivalently, on $n_h(\Rgs)$ where $\Rgs$ is the
Sun's galactocentric radius.  Though
we have made calculations explicitly for the nonuniform density
models, the local values are close to the reference model values. This
follows because on the Galactic scale most surveys probe regions near
the solar system.

\subsection{Disk, Spheroid, Halo and Globular Cluster Contributions}

The local column density of MSPs, $N_d \sim 50^{+30}_{-20} $ kpc$^{-2}$,
may be combined with the disk surface mass density ($\sim 66\pm 8
\msun$ pc$^{-2}$ for Oort K giants, \cite{b84}) to infer that the
number of MSPs per unit disk mass $(dN/dM)_{disk} \approx
7.6^{+4.4}_{-3.1} \times 10^{-7} \msun^{-1}$ (the range reflects only
the uncertainty in $N_d$). The total number of MSPs in the Galactic
disk scaled to the disk mass $M_{disk}$ is 
$3.0^{+1.8}_{-1.2} \times 10^4 (M_{disk}/
4 \times 10^{10} \msun)$ (for example, $M_{disk} = 3.7 \times 10^{10}
\msun$ [\cite{bs82}] by one estimate; $(3.5-4.6) \times 10^{10}
\msun$ [\cite{co81}] by another). The total does {\it not}
include a correction for beaming.

Estimates of the local spheroid mass density are uncertain, e.g.
$\rho_{sph} = 1.88 \times 10^{-4} \msun $pc$^{-3}$ (\cite{bss82}) or
$\rho_{sph} = (1.11-1.25) \times 10^{-3} \msun $pc$^{-3}$
(\cite{co81}). If $(dN/dM)_{disk} = (dN/dM)_{sph}$, then the
spheroid makes a contribution to the MSP number density $n_{sph} =
0.14$ kpc$^{-3}$ or $(0.84-0.95)$ kpc$^{-3}$, respectively. 
The upper limit we have derived for a 
uniform density model, $n_h \simless 0.42$ kpc$^{-3}$, is marginally
consistent. Future observations should be able to constrain 
contributions to the MSP population from Population II progenitors
more strongly.

Estimates of the dark matter halo density are
$\rho_{halo} = 9 \times 10^{-3} \msun$ pc$^{-3}$ (\cite{bss82}) or
$\rho_{halo} = (5.9-10.2) \times 10^{-3} \msun$ pc$^{-3}$ (\cite{co81}).
If the dark matter halo component satisfied
$(dN/dM)_{disk} = (dN/dM)_{halo}$ then its contribution is $n_{halo} =
6.8$ kpc$^{-3}$ or $(4.5-7.7)$ kpc$^{-3}$, respectively. Our limit on
$n_h$ implies $(dN/dM)_{halo}/(dN/dM)_{disk} < 0.06$ or $(0.05-0.09)$,
respectively.

Today's globular clusters are known to have a significant enhancement
of MSPs relative to the disk. With considerable uncertainty, Phinney
and Kulkarni (1994) estimate $(dN/dM)_{gc} \approx 50 (dN/dM)_{disk}$.
If half of the original globular cluster system has been destroyed
(e.g. a total mass $M \approx 5 \times 10^7 \msun$), if the MSP content was
similarly enhanced and if these MSPs orbit like the observed clusters,
then the contribution to the local mass density is $1.1 \times 10^{-3}
\msun$ kpc$^{-3}$ and the MSP number density is $4.2 \times 10^{-2}$
kpc$^{-2}$. Our limit on $n_h$ does not provide a strong constraint.

\subsection{Conclusions}

The observations place an upper limit on a diffuse halo-like
contribution to the MSP density that is roughly 1\% of the MSP disk density
at midplane.

\section{SPACE VELOCITIES OF MSPS}\label{sec:vel}

Our results indicate that millisecond pulsars are a low-velocity
population, at least when compared with young, high-field  pulsars.  
We have found
that the 3D rms velocity of MSPs in the galactic disk is $\sim 84$
\kms, about a factor of 5-7 lower than that of young, strong field
pulsars (Lyne \& Lorimer 1994; Cordes \& Chernoff 1996).  We have
reached this conclusion by determining the spatial distribution of
MSPs, by excluding the existence of  a significant 
non-disk population and by modeling
the motion of objects in the gravitational potential of the Galaxy.

\subsection{Comparison with Proper Motion Data}
\label{sec:pm}

We may compare
our results with direct measurements of proper motion using 
interferometric and pulse-timing methods;  the indirect method of
interstellar scintillation has also yielded determinations of 
MSP transverse speeds.  To date, there are timing proper motions
on eight MSPs: 
J0437-4715 (Bell \etal 1995),
B1257+12 (Wolszczan 1994),
J1713+0747 (Camilo, Foster \& Wolszczan 1994),
B1855+09 \& B1937+21 (Kaspi \etal 1994), 
B1957+20 (Arzoumanian, Fruchter \& Taylor 1994),
J2019+2425 and J2322+2057 (Nice \& Taylor 1995).
There are also scintillation speeds on some of these and other pulsars, 
B1855+09 (Dewey \etal 1988),
B1937+21 (Cordes \etal 1990), and 
J0437-4715, J1455-3330, J1730-2304 \& 2145-0750 (Nicastro \& Johnston 1995).
These MSPs have transverse speeds that are less than 100 \kms, except
for B1257+12, which has a speed of 285 \kms at its nominal distance
of 0.62 kpc and B1957+20, which has $\vperp \sim 173$ \kms at a distance of
1.2 kpc (Aldcroft, Romani \& Cordes 1992).  
For the most part, these objects are consistent with
our determination of the 3D rms velocity based on the locations of 22 MSPs
and the absence of MSPs in substantial portions of the volumes searched
in high-latitude surveys.   However, the estimated transverse speed
for B1257+12 is inconsistent with the overall distributions in $z$ and
velocity that we have derived, even though it was included in the fitting.
One possibility is that its distance is overestimated, perhaps by as much
as a factor of two, an amount sufficient to bring it into consistency with
the statistical distribution.  It is also possible that there are 
several evolutionary paths for producing MSPs (cf. \S\ref{sec:lmxb}),
most of which produce low-velocity MSPs with others creating rarer,
faster MSPs.

Further study of larger samples of MSP proper motions will result
from a combination of new surveys, which will discover large numbers
of MSPs (cf. \S\ref{sec:searches}), and use of timing and VLBI techniques.
Use of the VLBA in conjunction with the Arecibo telescope and the
Green Bank Telescope should allow measurement of proper motions
for dim and slow MSPs out to a few kpc.

\section{BIRTH RATES OF DISK \& HALO MSPS}\label{sec:birthrate}

For the disk-only model of \S\ref{sec:disk}, we have found the column
density of MSPs with $P_1 = 1.56$ ms, $L_{p1} = 1.1$ mJy kpc$^{-2}$
for a plane-parallel model in z to be $N_d \approx 50^{+30}_{-20}$
kpc$^{-3}$ (Table \ref{tab:bestfit}).  The implied number of MSPs in a
disk of radius $R_d$ with $P > P_1$ and $L_p > L_{p1}$ is
\begin{equation}
N_{MSP}(>P, >L_p)
    \approx 1.6^{+0.9}_{-0.6}\times 10^4 
         \left ( \frac{R_d}{10\, {\rm kpc}}\right )^2
	 \left( P \over 1.56 \,{\rm ms} \right)^{-1\pm0.33}
	 \left( L_p \over 1.1 \,{\rm mJy\,\, kpc}^{-2} \right)^{-1\pm0.2},
\label{eq:nmsp-unif}
\end{equation}
where the upper and lower values denote the 68\% interval. Extrapolation
on a per mass basis from the local disk surface density to  
a total disk mass, $M_{disk}$, implies
\begin{equation}
N_{MSP}(>P, >L_p) 
    \approx 3.0^{+1.8}_{-1.2}\times 10^4 
         \left ( M_{disk} \over 4 \times 10^{10} \msun \right )
	 \left( P \over 1.56 \,{\rm ms} \right)^{-1\pm0.33}
	 \left( L_p \over 1.1 \,{\rm mJy \,\, kpc}^{-2} \right)^{-1\pm0.2} .
\label{eq:nmsp-mass}
\end{equation}
These estimates do not include any correction for pulse beaming, whose
influence is highly uncertain for MSPs.  Estimates for this correction
range from 1 to 3 (e.g. Bailes \& Lorimer 1995). The totals are
sensitive to the cutoff at small periods and at small luminosities.

The corresponding birthrate for MSPs, if constant over
a galactic age $10^{10}$ yr, for the uniform disk
\begin{equation}
\dot N_{MSP}(> P_1) = 
    1.6^{+0.9}_{-0.6} \times 10^{-6} \,\, {\rm yr^{-1}}
         \left ( \frac{R_d}{10\, {\rm kpc}}\right )^2, 
\label{eq:birthrate}
\end{equation}
and for the extrapolated surface density is
\begin{equation}
\dot N_{MSP}(> P_1) = 
    3.0^{+1.8}_{-1.2}\times 10^{-6} \,\, {\rm yr^{-1}}
         \left ( M_{disk} \over 4 \times 10^{10} \msun \right ) .
\end{equation}

From our constraints on diffuse populations of MSPs, we conclude that,
in the vicinity of the Sun, the MSP birth rate per unit volume is
100 times less than that from the disk.

\subsection{Comparison with Other MSP Population Studies}

Our estimates may be compared with those derived by 
Bailes \& Lorimer (1995), Lorimer (1995)
and Lorimer \etal (1995), who used the Vivekenand \& Narayan
scale-factor method to determine the number of MSPs in the Galaxy and
the associated luminosity function.  In their analyses, specific spatial
distributions for the MSPs  were adopted to derive the scale factors.
BL assumed two different scale heights (0.3 and 0.6 kpc)
along with a fixed radial distribution to estimate $10^{4.4}$ and
$10^{4.6}$ MSPs, respectively, 
for $L_p > 2.5$ mJy kpc$^2$ and  if all MSPs beam toward us.

Lorimer \etal (1995) use the radial distribution of Lorimer \etal (1993)
(a Gaussian with radial scale of 4.8 kpc) and a Maxwellian velocity 
distribution with rms velocity = $\sqrt 3 \times 100$ km s$^{-1}$ to 
estimate $(1.3\pm0.2) \times 10^4$ MSPs in the Galaxy that are beamed
toward us with $L_p > 10$ mJy kpc$^{2}$.    

Lorimer (1995) deduced lower bounds on the scale height and mean 3D
space velocity for MSPs of 0.5 kpc and 80 km s$^{-1}$, respectively.
These bounds are consistent with our determinations.

The numbers of pulsars derived by BL and Lorimer \etal (1995) are
greater than the estimate in Eq.~\ref{eq:nmsp-mass} by a factor $\sim
2-3$ for a luminosity cutoff of $2.5$ mJy kpc$^2$. Since most of the
MSPs known are near the Sun (within 2 kpc) an extrapolation to the
whole Galaxy is necessary. The radial distributions used by BL and
Lorimer {\it et al.} effectively multiply the uniform disk model
result by $\sim 1.6$ and match our own extrapolation (based on scaling
up the local disk surface density to the given total disk mass in 
Eq.~\ref{eq:nmsp-mass}). The extrapolation introduces uncertainty but the
differences accrue from the following factors.  First, the z scale
height implied by the Lorimer \etal velocity distribution is larger
than that derived by us by about a factor of 2.  Second, our inclusion
of scintillation effects yields a search volume that is about 30\%
larger than otherwise.  Third, Lorimer \etal include four long period
pulsars in their analysis with $P > 295$ ms that we exclude from the
MSP sample.  Together these differences in the assumed spatial
distributions and MSP samples explain the size of the differences in
the estimated total number of MSPs in the Galaxy.

BL synthesized a luminosity function for MSPs after correcting the
observed numbers of pulsars for the volume scale factors.   Their
luminosity function is consistent with a power-law slope of $-2$
(according to our definition of ${f_L}_p$) but with a roll-off
below 10 mJy kpc$^2$.   Our method is able to constrain the lower
cutoff on the luminosity function because we evaluate our results at
values for $L_p$ other than those of actually detected pulsars. 

Similarly, BL suggest that the period distribution decreases
in going from 1 to 10 ms and roughly estimate  that there can be no more
than $10^{4.3}$ MSPs with periods with P = 1 ms.     
Our results suggest that the number of pulsars between 1 and 1.5 ms
is approximately 50\% of the number with $P > 1.5$ ms, or about
5000 pulsars. 

\section{RELATIONSHIP TO LOW-MASS X-RAY BINARIES}
\label{sec:lmxb}

\subsection{Scale Heights of LMXBs and MSPs}

The evolutionary paths that lead to MSPs are poorly understood (for
a review see \cite{b95}).  If
all MSPs are ``spun up'' by mass transfer from a companion star during
an LMXB phase, then the birth rate of LMXBs must exceed that of MSPs.
Kulkarni and Narayan (1988) estimated that the birthrate of field
LMXBs is about 1-10\% of the birthrate of field MSPs for an
assumed LMXB lifetime of $10^9$ years. With a diminished LMXB lifetime
($10^7$ years), the birthrates are brought into agreement. Our
improved estimate of the total number of MSPs in the Galaxy does
not significantly alter the rate mismatch nor its strong dependence on
LMXB lifetime. However, our work of the last section
does point out that the extrapolation
from the local MSP population to that of the Galaxy is uncertain by
a factor of $\sim2 - 4$ (and an additional factor of $1-3$ for
beaming).
As we will argue below, the kinematic similarity of the
LMXB and MSP population manifested in the observed scale heights is
reasonably strong evidence of an evolutionary link; given the great
uncertainty in LMXB lifetimes, the best evidence for a causal
connection between the two populations is not found in the relative
number but in the similar spatial distribution. In addition, a comparison of
the scale height distribution of MSPs with that of LMXBs can place
significant constraints on evolutionary scenarios leading to MSPs.

The galactic LMXB scale height derived from analysis of a flux-limited
sample (\cite{np93}) is (0.44-1.17) kpc.  Alternatively,
based on distance estimates to a subset of LMXBs, van Paradijs and
White (1995) infer a scale height of 0.5 kpc and, furthermore, argue that the
LMXBs are predominantly located at Galactocentric radii less than 5
kpc.  Although the two vertical scales are comparable, the
interpretations are quite different. Van Paradijs and White assume
that the LMXBs have Pop II progenitors
and that the scale height is set by large
velocity kicks at birth (of order 400 km s$^{-1}$) and the local disk
acceleration, which they argue is 2.5-4 larger in the relevant inner
regions of the Galaxy than locally. This analysis ignores the finite
birth scale height and the lifetime of the objects.  Naylor and
Podsiadlowski, on the other hand, infer that the LMXBs derive from Pop
I stars and have a scale height perpendicular to the plane that is roughly
like that of the observed thin disk (with a small additional kick),
which has a nearly constant value.

Our {\it local} determination of the MSP scale height is (0.53-0.81)
kpc, comparable with the above LMXB estimates. If MSPs are descendants
of the LMXB phase and if the scale height increases with age, then the
local MSP population should have a scale height greater than or equal
to the LMXB value. If the kicks were as large as suggested by Van
Paradijs and White, the minimum local scale height of the MSPs would
be (1.25-2) kpc, clearly inconsistent with our results.  In addition,
our upper limit on the diffuse number density of pulsars suggests
that the observed MSPs were born in the disk (Pop I). A consistent
interpretation of the LMXB and MSP data is that both are Pop I and
both are derived from a similar evolutionary channel.

\subsection{Origin of MSP Space Velocities}

In future work, we will discuss how the observed scale
height of MSPs and the inferred $z$ velocities ($\sim 50$ km s$^{-1}$) 
place stringent constraints on the evolution of binary systems
that lead to MSP formation.  One of
the main problems in understanding the LMXB evolution path is that the
formation rate ($10^{-6}$ yr$^{-1}$) in the Galaxy is so small that the pathway
is {\it a priori} special. A proposed scenario is as follows (\cite{wk94}).  A binary
composed of a massive star ($M_1 = 10-20 \msun$) and a light companion
($M_2 < 0.12 M_1$) with an initial orbital separation less than about
1000 $\rsun$ will pass through an epoch of unstable mass transfer and
common envelope evolution once the massive star begins to swell.  The
interaction ejects much of the envelope, drawing the pair to very
small distances of separation.  If the helium core of the primary is
sufficiently massive it is able to continue to burn and collapse even
after its outer hydrogen envelope has been removed. The
resultant neutron star has an orbit far smaller than
the size of the original giant primary, an essential requirement if an LMXB
phase is to take place. Starting with the pre-supernova system we have
analyzed how a tight binary is affected by a combination of (1)
asymmetric SN kick, (2) impact of ejected shell and (3) dissipative
processes in the eccentric, surviving binary.

Two effects sculpt the properties of the binaries that survive to give
LMXBs and/or MSPs. The intrinsic kick given a neutron star by the
supernova explosion unbinds loosely bound binaries while the impact of
the supernova shell on the secondary is responsible for destroying
and/or unbinding tight binaries. The surviving binaries occupy a
relatively narrow range in pre-supernova orbital separation, primary
and secondary masses and have a limited range of
center of mass velocities. Most of this analysis is independent of the
specific evolutionary pathway leading to the pre-SN progenitor. (We
present the details of this analysis in Chernoff \& Cordes 1996b.)

\section{SELECTION EFFECTS AGAINST FAST BINARIES}\label{sec:orbits}

Orbital motion causes MSPs in compact binaries to be missed in surveys
that assume the pulse period to be constant rather than Doppler
shifted (e.g. \cite{jk91}).  All blind surveys for MSPs, including the
8 analyzed in this paper, make this assumption.  One circumstance in
which the results of previous sections may be altered is if the spin
period and/or luminosity depend in some way on orbital period.  For
example, a relation between spin and orbital period might be expected
on general evolutionary grounds for spun up MSPs (\cite{arcs82},
\cite{rs83}).  Some observations suggest a weak positive correlation
(\cite{lzc95}) implying that the selection against detecting spin
periods less than 1.5 ms may be stronger than we have estimated.
Because the measured correlation is weak we believe that any
modification to the distribution of spin periods on this account will
be modest.  In any case, fast binaries have been missed in MSP surveys
and their ultimate detection can only increase our estimated space
densities for MSPs.

We now give a brief account of survey sensitivity to binary orbital
period.  The observation time $T\lesssim 1$ min for most of the
Arecibo surveys, so that orbital effects are negligible for $\porb
\gtrsim 1.\!^h6\,P^{-3/4}$ (P in ms) for WD companions
with $M_2 = 0.3\msun$.  However, surveys 4,7 \& 8 with $T\sim 3$ min
are insensitive to orbital motion only for $\porb \gtrsim
8.\!^h5\,P^{-3/4}$.  Weighted by volumes searched, the surveys with
longer $T$ contribute strongly to an overall selection against MSP
binaries with short periods.  Indeed, J0751+1807 with $\porb = 6.^h3$
was discovered in survey \# 4 in a single harmonic, the higher
harmonics having been attenuated by orbital motion (\cite{lzc95}).
That yet-faster binaries with fairly massive WD companions exist is
certain because objects like J0751+1807 experience orbital decay due
to gravitational radiation on less than a Hubble time.  Indeed, if
MSP-WD binaries achieve $\porb\lesssim 8^h$ solely due to such
inspiral, then it may be shown that the orbital period distribution
$dN/d\porb \propto \porb^{5/3}$.  Overall, our conclusions are
unaffected for MSPs in binaries with orbital periods $\simgreat 6$ hr
and with companion masses $\simless 0.3 \msun$.

Proper consideration of orbital effects --- and estimation of the MSP
orbital period distribution --- requires an analysis of search volumes
as a function of orbital period and  companion mass as well
as spin period and luminosity.  Such a study is beyond the scope of the
present paper.

Future surveys made with greater sensitivities than heretofore
using the upgraded Arecibo Observatory and the new GBT will be able to 
probe to greater distances while also circumventing orbital suppression
of Fourier harmonics.  Furthermore,  algorithms that correct
for orbital motion are becoming much more feasible with the prospect
of computers with teraflops capability.

\section{OPTIMAL SEARCHES FOR MILLISECOND PULSARS}
\label{sec:searches}

The population distributions we have derived may be used to optimize
new searches for MSPs.  Search sensitivities (Appendix \ref{app:smin})
depend on sky background, dispersion and scattering as well as on the
flux densities and periods of the MSPs.  Consequently, the optimal
search is frequency and telescope dependent.  Here we illustrate the
contributions from different effects by showing $\delta V_{s}$, the {\it
search volume} (the volume searched in a beam area,
averaged over $P$ and $L_p$, Eq.~\ref{eq:volsea}) 
and the {\it detection volume}, $\delta V_d$, 
(the volume searched
weighted by the dimensionless density of MSPs, 
Eq.~\ref{eq:dvolume}).  To calculate each we use the the best-fit
distributions for $P$ and $L_p$ for the exponential disk model of
\S\ref{sec:disk} and Table~\ref{tab:bestfit}.  By definition
$\delta V_d \le \delta V_{s}$: 
surveys that search much more deeply than the
scale height of the MSP population yield  $\delta V_d \ll \delta V_{s}$.  For
concreteness, we consider a survey conducted by telescopes like the
under-construction Green Bank Telescope (GBT) and a hypothetical
analog in the Southern hemisphere, in order that we may consider a
full sky survey.  We assume receiver and survey parameters such that
the minimum detectable flux density is about 2 mJy when looking at
high galactic latitudes and long periods.

\begin{figure}
\plotfiddle{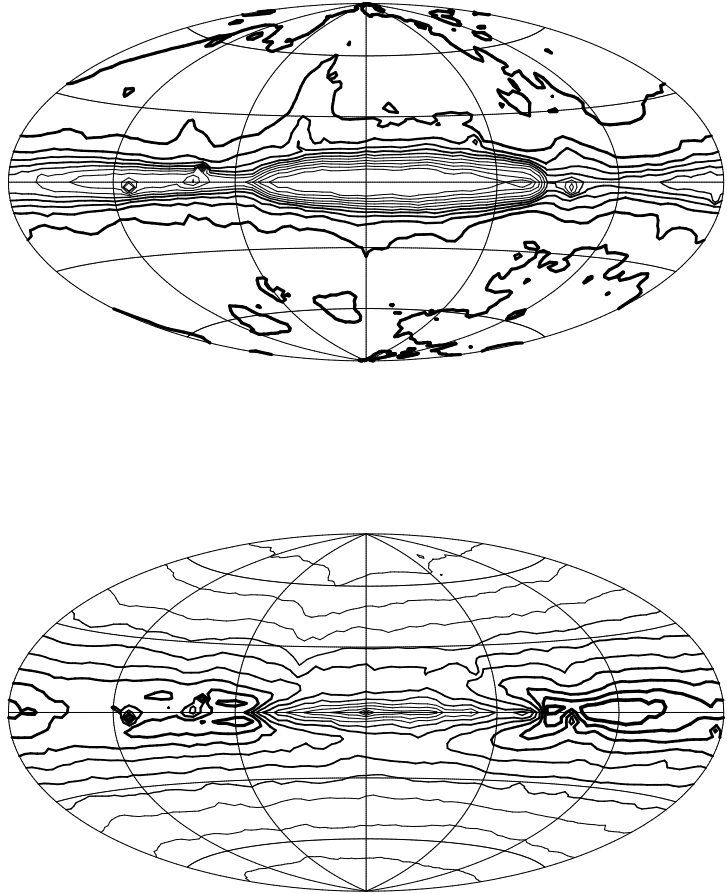}{5.5truein}{0.}{80}{80}{-244}{-172}
\caption{(Top:) Aitoff projection in galactic coordinates
showing the \protect{\it search volume} (kpc$^{3}$ deg$^{-2}$)
for a hypothetical full-sky survey at 430 MHz.  The calculated volumes
are averages over the period and luminosity distributions in our best-fit,
exponential disk model.
The minimum volume is toward the galactic center.
The thinnest contour (toward the inner Galaxy) corresponds to
the least volume ($10^{-3.6}$ kpc$^3$ deg$^{-2}$) while the thickest line
(at high latitudes) corresponds to the
greatest volume ($\sim 10^{-1.9}$ kpc\protect$^{3}$ deg$^{-2}$).
Most of the variation is from a deep minimum toward the Galactic center
to a shallower variation beginning at $\vert b \vert \sim 15^{\circ}$.
The greatest volumes searched are those toward the highest latitudes.
Structure is seen in the plotted contours (e.g. the North Polar Spur)
because the search
sensitivity and, hence, depth are dependent on the sky background and on
dispersion and scattering.
(Bottom:) Similar projection for the
\protect{\it survey detection volume}.
The minimum contour (toward the Galactic center) is
$10^{-4.3}$ kpc$^3$ deg$^{-2}$ while the maximum (thickest line) is
$10^{-2.8}$ kpc$^3$ deg$^{-2}$ at
latitudes $\vert b \vert \sim 5^{\circ}$ and
longitudes $\vert\ell\vert \simgreat 50^{\circ}$.  Note that there is no
Galactocentric radial dependence of our assumed MSP number density,
so all the structure at $b = 0^{\circ}$ (and other lines of constant
latitude) is due to the depth
of the survey.
}
\label{fig:aitoff}
\end{figure}

Figure \ref{fig:aitoff} shows $\delta V_{s}$ and $\delta V_d$ per square
degree for a search at 430 MHz.  Search volumes (top portion of
figure) increase more or less monotonically with galactic latitude but
level off for $\vert b \vert > 30^{\circ}$.  The volume is smallest
toward the galactic center where the sky background is high and
dispersion and scattering effects are large.  By contrast, the 
detection volume (bottom part of figure) is maximum for $\vert b \vert \sim
20^{\circ}$ and $\vert l \vert \simgreat 50^{\circ}$ and corresponds to
directions that allow the largest $\int_0^{D_{max}} dD D^2
n_p(D,\ell,b)$. The latitude constraint ensures that the search depth
does not exceed the MSP scale height. The longitude restriction
follows from the variation of the search volume in the plane.  The
detailed shapes of the contours are dependent on the survey frequency
and duration (per direction) but suggest that future surveys which
concentrate on low latitudes will maximize the number of new
discoveries.  However, deep high latitude surveys will better
constrain the falloff toward larger $z$ of the disk population as well
as place tighter constraints on or make detections of any {\it bona
fide} diffuse or halo-population pulsars.
For the hypothetical survey depicted in Figure \ref{fig:aitoff},
a total detection volume $\sim 13.3$ kpc$^{3}$ is sampled,
corresponding to discovery of $\sim 585^{+330}_{-210}$ disk MSPs.
This number is for a uniform disk component; any galactocentric
radial dependence, likely to increase the number of MSPs toward the
inner Galaxy, will only increase the number of detected MSPs.

\section{DISCOVERING FAST PULSARS}\label{sec:fast}

Our fitting indicates that available survey data already place useful
constraints on the minimum spin period in the MSP population (cf.
\S\ref{sec:disk} and Table~\ref{tab:bestfit}).  The reason such
constraints may be placed is found in Figure \ref{fig:surveys} which
shows that, for periods less than 1 ms, a nonzero (though small)
volume has been searched.  Here we estimate how much additional volume
must be searched in order to expect to find pulsars with $P_1 < P <
P_{\rm fast}$ where $P_{\rm fast}$ is the maximum period of interest.

Using Eq.~\ref{eq:npave} we derive an upper bound on the volume that must
be searched (evaluated at the minimum MSP period, $P_1$), in order that we find
pulsars with periods faster than $P_{\rm fast}$.  Since surveys at low periods
do not see to large $D_{max}$, we assume that they do not see as 
far as the $z$ scale height $\sim 0.5$ kpc.  Performing the integrals we
may write
\begin{equation}
\npave \simgreat n_d \left (1 - P_1/P_{\rm fast} \right )
         \left [
             \frac{1}{3}\Omega_b 
             \langle L_p^{3/2} \rangle 
             S_{\rm min}^{-3/2}(P_1) 
         \right ].
\label{eq:nplimit}
\end{equation}
Defining the term in square brackets as $V_S(P_1)$ and requiring
$\npave \sim 1$ to obtain a likely detection, we find an {\it upper bound} on
the required search volume to be
\begin{equation}
V_S(P_1) = \frac{1}{n_d (1 - P_1/P_{\rm fast})}.
\label{eq:vslimit}
\end{equation}

Evaluating Eq.~\ref{eq:vslimit} for 
$P_{\rm fast} = 1.5$ ms and
$n_d \sim 44$ kpc$^{-3}$,
we find that, as a function of the minimum period, $V_S(P_1)$ ranges
from $\sim 1/25$ kpc$^3$ for $P_1 = 0.65$ ms (our 99\% lower bound on $P_1$)
to $\sim 1/3$ kpc$^3$ for $P_1 = 1.4$ ms.  Comparison with 
Figure ~\ref{fig:surveys} shows that, for the luminosity assumed for
that figure ($L_p = 16$ mJy kpc$^2$), the Parkes survey (\#7) yields
search volumes at these $P_1$ that are comparable to those needed
to yield a detection  of a pulsar faster than 1.5 ms.  However, 
the assumed $L_p$ for the figure is larger than average and the Parkes survey
observes to depths that, for some directions, exceed the $z$ scale height.
Consequently, it is not surprising that a pulsar faster than 
PSR B1937+21 (P = 1.56 ms) has not been found.   Nonetheless, future surveys
should be able to probe this region of period space and either find
fast pulsars or determine better the period cutoff to the MSP population. 
Our results indicate that deeper surveys at low galactic latitudes
(e.g. $\vert b \vert \simless 10^{\circ}$) will yield the search volume
needed to accomplish these goals. 

\section{DISCUSSION}
\label{sec:discussion}

Through a likelihood analysis, we have constrained the period and
pseudo-luminosity distributions to be steep power laws with
slopes $\sim -2$.  The distributions imply that the population of MSPs
increases rapidly to smaller periods and smaller pseudo-luminosities.
We infer a minimum period $P_{min} > 0.65$ ms at 99\% confidence and a
minimum luminosity cutoff $L_{p1} = 1.1^{+0.4}_{-0.5}$ mJy kpc$^2$. The column
density of MSPs in the local vicinity of the solar system is $N_d \sim
50^{+30}_{-20}$ kpc$^{-2}$.  The limits on a diffuse halo-like
component are $\simless 1$\% of the midplane density. All these results are
essentially identical for each of the models we have analyzed.
Estimates of the total number of MSPs in the Galaxy are uncertain.
Extrapolating on a per mass basis from the local disk surface density
to a total disk mass, $M_{disk}$, we find $N_{MSP} \approx
3.0^{+1.8}_{-1.2}\times10^4\left(M_{disk}/4\times10^{10}\msun\right )$ 
for cutoff period $1.56$ ms and cutoff luminosity 
$1.1$ mJy kpc$^{-2}$ (without correction for beaming).

Our analysis assumes specific forms for the period and luminosity
distributions, namely power-law functions, that undoubtedly influence
the specific values for numbers of pulsars in a given period range
and also on the minimum period cutoff.   We have not tested other 
mathematical forms for these distributions, so the true cutoff for the
period distribution may be different than we have derived.  Nonetheless,
because the period distribution montonically increases with decreasing
period, our quoted minimum period is {\it larger} than it would be
for a function that plateaus or decreases with decreasing period  below
1.56 ms.   We consider the most important implication of our derived
minimum period to be that MSPs faster than those already found
may indeed  be present in the Galaxy:  the surveys done heretofore
cannot rule out their existence.  In addition, modeling of
the spinup process using full general relativity (Cook \etal 1994a,b)
implies that gravitational instabilities do not prohibit the formation
of very fast MSPs.   Of course the ultimate existence proof for MSPs
with $P<1.56$ ms lies in future surveys that can explore large
volumes of the Galaxy at these small periods.   Such surveys will 
be feasible with new spectrometers that can sample more frequency
channels at faster rates and with post processing that can contend
with  motion of fast pulsars in binaries. 

Another implication of our results on the period distribution is that,
if MSPs exist due to accretion-driven spinup of neutron stars, 
then accretion must ensue for sufficiently long times that 
periods shorter than 1.56 ms can be achieved.   From the work of
Cook \etal (1994a,b), such accretion appears possible without
requiring typical ages for LMXBs that are so long as to 
resurrect the discrepancy between birth rates for MSPs and LMXBs. 

The observed scale height of MSPs implies that they are a low-velocity
population among neutron stars, having an rms speed that is about a
factor of 5 smaller than that of young pulsars with much stronger
magnetic fields. A part of the total inferred dispersion we attribute
to a kick unique to the evolution of MSP systems ($\sim 40$ km
s$^{-1}$) and the rest to the effect of diffusive processes that
increase the dispersion of old objects.  A number of kinematic
signatures that should be evident in larger MSP samples (transverse
motions, asymmetric drift, shape of velocity ellipsoid) are described.

The disparity in velocity between the low-field MSPs and high-field
pulsars might be taken as evidence that the two empirical classes of
neutron stars are born through substantially different processes. If
MSPs are produced largely through accretion-induced collapse of a
white dwarf and if that process yields only a small kick to the
resultant NS compared to Type II supernova then the observed
dispersions of MSPs and high-field pulsars may find a natural
explanation. In any case, the similarity in scale height of MSPs and
LMXBs shows that the formation of the NS in both objects is
accomplished without substantial center-of-mass impulses and supports
the notion of an evolutionary connection.  On the other hand, if
binary survival after the type II supernova is the most significant
bottleneck in the production of LMXBs and their MSP descendants, it is
possible that the processes that dictate survival of the binary
system are also responsible for allowing only a limited range of
center-of-mass velocities.   Correlations between spin and orbital
periods and space velocity, such as those suggested by Bailes \etal (1994),
depend critically on the details of mass transfer and on the number of
evolutionary paths that lead to MSP formation.  
Elsewhere, we will present our detailed
analysis of the effects that sculpt binary survival.

We thank Z. Arzoumanian, M. Bailes, G. Cook, D. Lorimer, T. Loredo, S. Lundgren
and I. Wasserman for useful discussions.  This work was supported by
NSF Grants AST-91-19475, AST-95-30397, AST-92-18075 and NASA Grants
2581 and 2224 to Cornell University.  It was also supported by the
National Astronomy and Ionosphere Center, which operates the Arecibo
Observatory under a cooperative agreement with the NSF.

\vfill
\eject
\appendix
\centerline{\bf APPENDICES}

\section{SEARCH SENSITIVITIES}\label{app:smin}

The pulsar searches we consider involve the removal of dispersion
delays between the outputs of a multichannel receive using trial values
for the dispersion measure.  The resultant time series is then
Fourier analyzed.
Suppose an $\nfft$-length Fast Fourier Transform (FFT) 
is calculated from the time series for each 
trial dispersion measure.  With a sample time $\Delta t$ and pulse period P,
harmonics appear in frequency bins
\begin{equation}
k_{\ell} = \frac{\ell\Delta t\nfft}{P},\quad \ell = 0,1,\ldots
\label{eq:kl}
\end{equation}
including a ``DC'' term ($\ell=0$) and the fundamental ($\ell = 1$).
Let the intrinsic pulse shape be $s(t), \,0\le t \le P$ so that a (short)
discrete Fourier transform (DFT) of this shape over a single pulse period is
$\tilde s({\ell}), \,\, \ell = 1, \ldots, M$.   
This function would determine the envelope of harmonic amplitudes in the long
FFT were it not for
additional contributions that derive from the post-detection
averaging time (or ``time constant''), from dispersion smearing across
individual frequency channels, and from pulse broadening due to
interstellar scattering (for distant sources).  In many surveys, post-detection smoothing is simply an RC filter
whose time domain response is a one-sided exponential function.    Interstellar
scattering produces nearly the same kind of time response, while
the dispersion time function is dictated by the shapes of receiver filters,
usually approximately Gaussian in form.    Letting the M-point DFTs of
the time constant, dispersion, and scattering functions be
$\tilde s_{tc}$, $\tilde s_{d}$ and $\tilde s_{s}$, respectively,
we may write the effective envelope function of harmonics  as 
\begin{equation}
\tilde s_{eff}(\ell) = 
\tilde s(\ell)
\tilde s_{tc}(\ell)
\tilde s_{d}(\ell)
\tilde s_{s}(\ell).
\label{eq:envelope}
\end{equation}
It is useful to define 
the ratio of the $\ell$-th harmonic to the DC value as
\begin{equation}
R_{\ell} \equiv \left |\frac{\tilde s_{eff}(\ell)} {\tilde s_{eff}(0)}\right |.
\label{eq:ratio}
\end{equation} 

Survey FFTs are analyzed by constructing partial sums of harmonics
(of the FFT magnitude or squared magnitude)
for different trial periods.   These sums are typically of
$N_h = 1,2,3,4,8$ and 16 harmonics, though there are variations on this. 
Suppose that a threshold $\eta_T$ is chosen that represents the number
of standard deviations in the FFT's magnitude.  This is typically 
$\eta_T \sim 6$ to 9 in order to minimize false-alarms when testing 
large numbers of spectral values (typically multiples of $10^9$) 
in a survey.   

The minimum detectable flux density for a sum of
harmonics $1,\ldots,N_h$ is
\begin{equation}
S_{min, N_h} = 
    \frac{\eta_T T_{sys}} {G\sqrt{N_{pol}\Delta\nu\Delta t\nfft}}
    \left (\frac {\sqrt{N_h}}{\displaystyle  \sum_{\ell=1}^{N_h} R_{\ell}} \right ),
\label{eq:smin_nh}
\end{equation}      
(where $T_{sys}$ is the system temperature [K];
       G is the telescope gain [K Jy$^{-1}$];
       $N_{pol}=2$ is the number of independent polarization channels
         included;
       $\Delta\nu$ is the total bandwidth; and
       $\Delta t$ is the sample interval).
For searches that analyze ${\rm \vert FFT \vert^2}$ 
rather than $\vert {\rm FFT} \vert$,
$R_{\ell} \to R_{\ell}^2$.
The actual minimum flux density 
depends on the number of harmonics that contribute significantly which,
in turn, depends on the duty cycle of the pulse. 
Because extrinsic effects (viz. dispersion and scattering),  
broaden the pulse, the optimal $N_h$ and corresponding 
$S_{min}$ are  strongly dependent on the 
observation frequency, distance, direction and  pulse period.
The direction dependence is manifested in the dispersion measure to which
a pulsar of given period may be detected.   Consequently $S_{min}$ is
the minimum over all $N_h$ considered in the analysis and 
may be written with dependences
\begin{equation}
S_{min} = S_{min}(\ell, b, P, DM, \nu, \Delta\nu, N_{ch}, T_{sys}, G, \ldots).
\label{eq:sminapp}
\end{equation}
In practice, surveys usually test only a subset of all possible harmonic
sums.  We take this into account when computing $S_{min}$ for each
survey. 

Note that our expression for the minimum flux density differs from that
often quoted in the literature (e.g. \cite{cnt96}),
which replaces the factor in large brackets in Eq.~\ref{eq:smin_nh}
with a factor $\sqrt{w/(P-w)}$, where $w$ is the pulse
width. The divergence of this factor as $w\to P$ is equivalent 
to {\it assuming}  $R_1 = 0$, which overestimates
the true $S_{min}$ because,  even when pulse smearing exceeds a pulse
period, the variable flux remaining at the fundamental frequency
can still be detectable for a luminous pulsar.   
Our expression takes this possibility into account, which corresponds
to $0 < R_1 \ll 1$.  

It is important to calculate accurately 
the minimum detectable flux density because it determines the galactic volume
searched.  This volume is small but not zero for very short periods
$\simless 1.5$ ms.

\section{INTERSTELLAR SCINTILLATIONS}\label{app:iss}

\begin{figure}
\plotfiddle{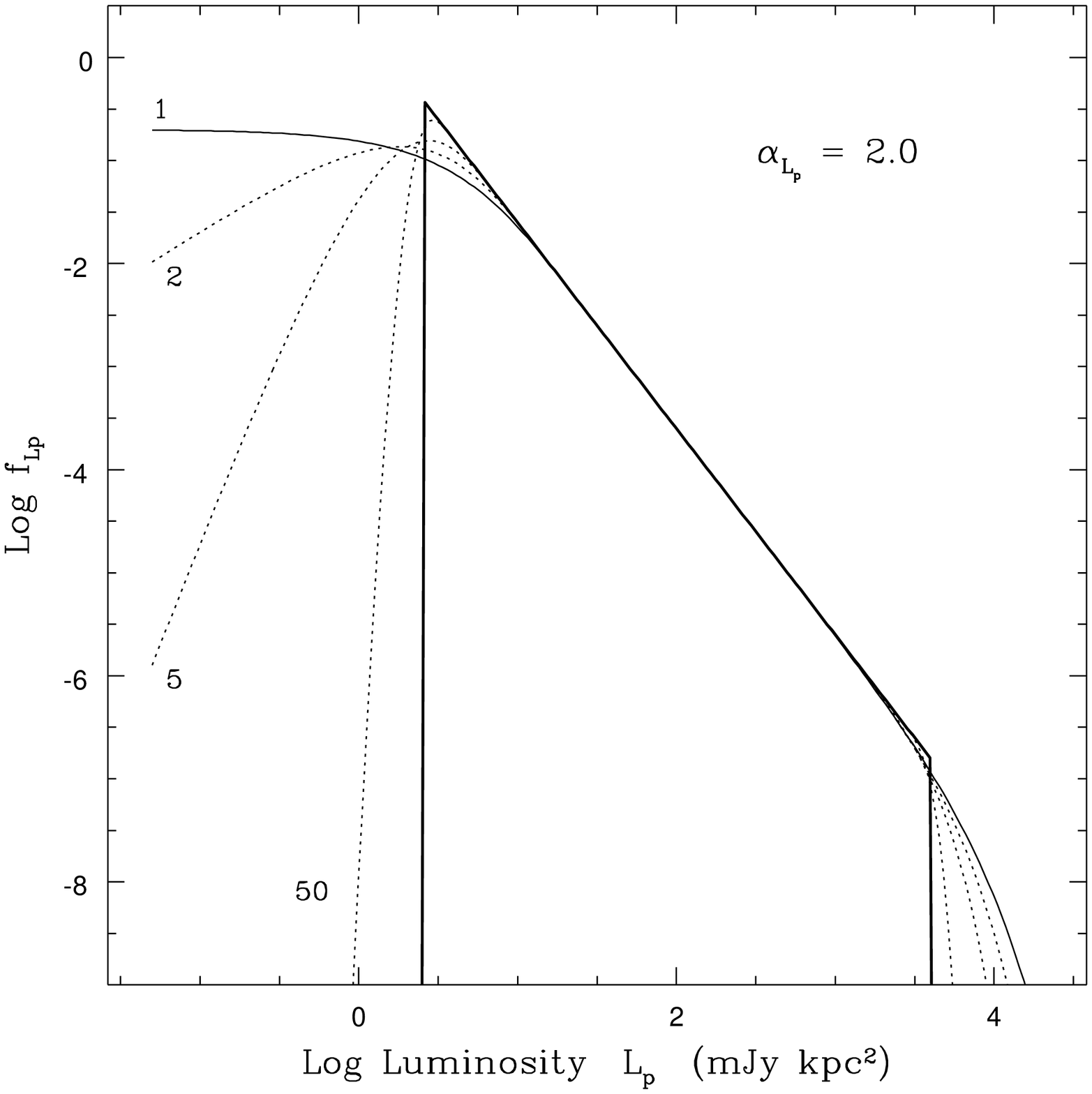}{5.0truein}{0.}{50}{50}{-144}{-72}
\figurenum{B1}
\caption{Luminosity functions with and without the effects of scintillations
included. (Heavy Solid Line:) Intrinsic luminosity function having a power-law
slope of $-2$.  (Light Solid Line:) The scintillated luminosity function when
interstellar scintillations are saturated and unquenched by time-bandwidth
averaging.  (Dotted Lines:) Scintillated luminosity functions with various
degrees of averaging, indicated by $\niss$ (cf. Eq.~\protect\ref{eq:splf}).
}
\label{fig:pdflp_iss}
\end{figure}
Interstellar scintillations are intensity variations in both 
time and frequency caused by multipath propagation through 
ionized gas.   At 400 MHz, both diffractive (DISS) and
refractive (RISS) interstellar scintillations contribute to the flux
variations of pulsars.   Here we restrict the discussion to DISS, which
will dominate RISS at 400 MHz and is especially important because its
probability density is skewed whereas RISS is symmetric.

DISS causes the pulsar flux density $S$ to vary as $\sprime = gS$,
where $g$ is the DISS gain that has a one-sided exponential distribution
when DISS is saturated and not quenched by time-bandwidth averaging
(Rickett 1990; Cordes \& Lazio 1991).   Except for the very nearest
pulsars ($D < 100$ pc), DISS is saturated at 400 MHz.  The characteristic
time and frequency scales of DISS diminish with increasing distance.  Use of
 finite bandwidth $B$ and data-span length $T$ will average over distinct
scintillation maxima,   increasing  the number of degrees of
freedom from 2 (for unquenched DISS) to 2$\niss$, where
\begin{equation}
\niss \sim 
       \left ( 1 + 0.2\frac{B}{\Delta\nu_d} \right )
        \left ( 1 + 0.2\frac{T}{\Delta\nu_t} \right ),
\label{eq:niss}
\end{equation} 
$\Delta\nu_d$ is the characteristic bandwidth of DISS, and
$\Delta t_d$ is the characteristic time scale (Cordes 1986).  
The characteristic bandwidth and time scale have been measured
for many pulsars and were used as input to the Taylor \& Cordes
(1993) model for pulsar distances.  For our purposes, we use the TC
model's estimation of the scattering measure along with the distance
and frequency to estimate the scintillation parameters.

The pdf of $g$ is
\begin{equation}
f_g(g, \niss) = 
   \frac{(g\niss)^{\niss}}{g\Gamma\left(\niss\right )} e^{-g\niss} U(g), 
\label{eq:fg}
\end{equation}
with $U(g)$ the Heaviside function  and $\Gamma$ is the gamma function.
As $\niss\to \infty$ (i.e. pulsars at large distances or observed at low
frequencies), $f_g$ tends toward a delta function, $\delta(g-1)$. 
  
We include scintillations in our analysis by defining a
scintillated pseudo luminosity, $\lpprime = gL_p$.
For a luminosity function $f_{L_p}(L_p)$, the corresponding
scintillated luminosity function is
\begin{equation}
f_{\lpprime}(\lpprime) = \int dg\, g^{-1} f_g(g, \niss) f_{L_p}(\lpprime/g).
\label{eq:splf}
\end{equation}

The distinctive effect of DISS is that, if the intrinsic luminosity function
has cutoffs at low and high luminosities, the scintillated luminosity function
will not.  In fact, the scintillated luminosity function will extend to
zero luminosity because the most probable scintillation gain 
(for $\niss = 1$) is zero.    Luminosities larger than the upper cutoff
will be seen owing to the long exponential tail of $f_g$ 
(again for $\niss = 1$).  Figure~\ref{fig:pdflp_iss} shows examples
of scintillated luminosity functions for several values of
$\niss$.  As $\niss\to\infty$, the scintillated luminosity function tends
toward the original, unscintillated luminosity function. 

In surveys where DISS is saturated and unquenched ($n_{ISS} = 1$)
and single trials are made on each sky position, the volume surveyed
is effectively increased by a factor
$\langle g^{3/2} \rangle = \Gamma(\frac{5}{2}) \sim 1.33$.
Multiple trials can increase or decrease this volume factor, depending
on how the results of the various trials are combined.

\begin{table}
\caption{\hfil MSP Survey Parameters}
\label{tab:surveys}
\begin{center}
\begin{tabular}{ccrrrrrr}
\tableline
Survey & Site & $\nu$    & $\Omega_S$ & $N_{MSP}$ & S$_{\rm sys}$ & $S_{\rm min_0}$ & Ref \\
       &      & (GHz) & (deg$^2$)  &           & (Jy)    & (mJy) \\
(1)    &  (2) &  (3)     &   (4)      &  (5)      &    (6)        &   (7)  &  (8) \\
\tableline
\tableline
\\
1  &  A  &  0.43  &  680  &  3  & 3   &  0.5  &  1 \\
2  &  A  &  0.43  &  235  &  4    & 3   &  1.0  &  2 \\
3  &  A  &  0.43  &  250  &  0    & 3   &  0.4  &  3 \\
4  &  A  &  0.43  &  7    &  1    & 3   &  0.2  &  4 \\
5  &  A  &  0.43  &  682  &  2    & 3   &  0.7  &  5 \\
\\
6  &  A  &  0.43  &  150  &  1    & 3   &  0.4  &  6 \\
7  &  P  &  0.44  & 20,600 & 10   & 90  &  3.0  &  7 \\
8  &  J  &  0.41  & 1,650  & 1    & 70  &  3.1  &  8 \\ 
\\
\tableline
\end{tabular}
\end{center}
\noindent
Sites: A = Arecibo, J = Jodrell Bank, P = Parkes.

\noindent
References: (1) Camilo et al. 1996; 
            (2) Nice et al. 1995;
            (3) Thorsett et al. 1993; 
            (4) Lundgren et al. 1995;
            (5) Foster et al. 1995;
            (6) Wolszczan 1990;
            (7) Manchester et al. 1996;
            (8) Nicastro et al. 1995.
\end{table}

\begin{table}
\caption{\hfil Millisecond Pulsars Used}
\label{tab:msplist}
\begin{center}
\begin{tabular}{lrrrrrrrrr}
\tableline
MSP Name & $\ell$  &  b  &  P  &  log $\Delta P$  &  S    &  $\Delta S$  &  
         $D_L$  & $D_U$ &   Ref  \\
         &  (deg)  & (deg) & (ms) &  (ms)        & (mJy) &    (mJy)     &      
         (kpc)  & (kpc) & \\
(1)      &  (2)    & (3) & (4) &  (5)             & (6)   &  (7)         &
(8)             &  (9)  &  (10) \\
\tableline
\tableline
\\
J0034$-$0534 & 111.5   &  $-$68.1    & 1.88  & $-$10.7  & 16 & 5 & 0.74  & 1.23 &  7  \\
J0437$-$4715 & 253.4   &  $-$42.0  & 5.76  & $-$11.4  & 600& 180 & 0.105 & 0.175 &  7 \\
J0613$-$0200 & 210.4 &   $-$9.3 & 2.19 &  $-$11.4 & 21 & 6 & 1.64 & 2.74 &
  7 \\
J0711$-$6830 & 279.5   &  $-$23.4  & 5.49  &   $-$4.1&  7 & 2 & 0.77 & 1.29 &
  7 \\
J0751+1807 & 202.7 &   21.1  & 3.48  &  $-$11.0 & 10 & 3 & 1.51 & 2.53 &
  4 \\
\\
J1012+5307 & 160.3 &  50.9   & 5.26  &  $-$10.7 & 30 & 9 & 0.39 & 0.65 &
  8 \\
J1045$-$4509 & 280.9   &  12.3   & 7.47  &  $-$10.7 & 20 & 6 & 2.43 & 4.05 &
  7 \\
B1257+12   & 311.3 &  75.4   & 6.22  &  $-$12.7 & 20 & 6 & 0.47 & 0.77 &
  6 \\
J1455$-$3330 & 330.7 &  22.6   & 7.99  &  $-$10.2 & 13 & 4  & 0.56 & 0.93 &
  7 \\
J1640+2224 &  41.1 &  38.3  & 3.15  &  $-$12.3 & 12 & 4 & 0.88 & 1.48 &
  5 \\
\\
J1643$-$1224 &   5.7   &  21.2   & 4.62  &  $-$10.5 & 75 & 23 & 4.84 & $\infty$ &
  7 \\
J1713+0747 &  28.8   &  25.2   & 4.57  &  $-$12.1 & 36 & 10 & 0.8   & 1.6 &
  5 \\
J1730$-$2304 &   3.1 &   6.0   & 8.12  &  $-$10.5 & 43 & 13 & 0.38  & 0.64 &
  7 \\
B1855+09   &  42.3   &   3.1   & 5.36  &  $-$12.5 & 31 &  9 & 0.70  & 1.30 &
  2 \\
B1937+21   &  57.5 &  $-$0.3   & 1.56  &  $-$12.7 &240 & 72 & 3.60  & 15.7 &
  2 \\
\\
B1957+20    & 59.2  &  $-$4.7    & 1.61  &  $-$12.52 & 20  & 6 & 1.15 &1.91 &   2 \\ 
J2019+2425 &  64.7  & $-$6.6 &  3.93  & $-$12.7 & 15  &  5.0   & 0.68 & 1.14 &
  2 \\
J2124$-$3358 &  10.9  & $-$45.4 & 4.93  & $-$10.2 & 20  &  6  & 0.18 & 0.30 &
  7 \\
J2145$-$0750 & 47.8  & $-$42.1 & 16.05 & $-$9.7 & 50 & 15 & 0.38 & 0.62 &
  7 \\
J2317+1439 & 91.4   & $-$42.4 &  3.45 & $-$12.7 & 14 & 5 & 1.4  & 2.3  &
  1 \\
\\
J2322+2057 & 96.5   & $-$37.3 &  4.81 & $-$12.6 & 4 & 2 & 0.5  & 1.1  &
  1 \\
J2229+2643 & 87.7   & $-$26.3 & 2.98 & $-$12.1 & 18 & 5 & 1.0 & 2.0 &
  1 \\
\\
\tableline
\end{tabular}
\end{center}
\noindent
 
\noindent
References: (1) Camilo et al. 1996;
            (2) Nice et al. 1995;
            (3) Foster et al. 1995;
            (4) Lundgren et al. 1995;
            (5) Thorsett et al. 1993;
            (6) Wolszczan 1990;
            (7) Manchester et al. 1996;
            (8) Nicastro et al. 1995.
\end{table}

\begin{table}
\caption{\hfil Best-fit Disk Models}
\label{tab:bestfit}
\begin{center}
\begin{tabular}{cllll}
\tableline
\\
Parameter &Gaussian & Exponential & Gaussian \\ 
          & in z    & in z        & in V     &units \\
\\
\tableline
\tableline
\\
$\alpha_P$     &  2.0$\pm$ 0.33  & 2.0$\pm$0.33  &  2.0$\pm 0.33$                   & --- \\ 
\\
$\alpha_{L_p}$ &  2.0$\pm$ 0.2   & 2.0$\pm$0.2   &  2.1$\pm 0.2$                    & --- \\ 
\\
${\sigma_z}$   & $0.65^{+0.16}_{-0.12}$   & $0.50^{+0.19}_{-0.13}$& ---             & kpc  \\
\\
${\sigma_{z,birth}}^{\dagger}$   &  --- &  --- & $0.1$  & kpc \\
\\
${\sigma_V}$   &  ---                     &  ---                  & $52^{+17}_{-11}$  & km s$^{-1}$ \\
\\
$n_d$          & $29^{+17}_{-11}$          & $44^{+25}_{-16}$ & $53^{+28}_{-18}$       & kpc$^{-3}$  \\
\\
$N_d$          & $52^{+29}_{-19}$          & $49^{+27}_{-18}$ & $49^{+27}_{-17}$  & kpc$^{-2}$  \\
\\
$P_1$            & $>1.0$  (95\%)         & $>1.0$  (95\%) & $>1.0$ (95\%) & ms      \\
                 & $>0.65$ (99\%)         & $>0.65$ (99\%) & $>0.70$ (99\%) &         \\
\\
${L_p}_1$        & $1.1^{+0.4}_{-0.5}$    & $1.1^{+0.4}_{-0.5}$& $1.1^{+0.4}_{-0.5}$ &  mJy kpc$^{2}$ \\
\\
\tableline
\end{tabular}
\end{center}
Confidence intervals, except where noted, are two-sided 68\% intervals.
$^{\dagger}$ fixed parameter.
\end{table}

\def\nhhat{{\hat n_h}}
\def\ndhat{{\hat n_d}}
\def\kms{{km s$^{-1}$}}
\def\kpc3{{kpc$^{-3}$}}
\begin{table}
\caption{\hfil Disk + Diffuse Models}
\label{tab:disk_halo}
\begin{center}
\begin{tabular}{lcccccc}
\tableline
\\
	& \multicolumn{2}{c} {Diffuse only}   &&  
    \multicolumn{3}{c} {Disk $+$ Diffuse}  \\
\cline{2-3}                   \cline{5-7} \\
Model & $\nhhat$ & $\displaystyle\frac{\cl(0,\nhhat)}{\cl(\ndhat,0)}$ && 
                   $\ndhat$ & $\nhhat$ & $\nhhat(\Rgs)$ \\
            & (\kpc3)   &  &  & (\kpc3) & (\kpc3) & (\kpc3)      \\ 
\tableline
\tableline
\\
uniform        & 1.5  & $10^{-21.3}$ &&  38& $<0.42$ (90\%) & ..  \\
density        &      &              &&    & $<0.84$ (99\%) & ..  \\ 
\\
\tableline
\\
$r_h = 1$ kpc & 4080 & $10^{-16.4}$  &&  38 & $< 1520 $ (90\%) & $<0.83$ \\
$s_h = 7/2$   &      &               &&     & $< 3040 $ (99\%) & $<1.66$ \\
\\
\tableline
\\
$r_h = 5$ kpc & 10.8 & $10^{-15.1}$  &&  38   & $< 3.3 $ (90\%) & $<0.31$ \\
$s_h = 2$   &      &                 &&       & $< 6.7 $ (99\%) & $<0.62$ \\
\\
\tableline
\\
\end{tabular}
\end{center}
\end{table}

\begin{table}
\caption{\hfil Velocity Moments for Nearby Pulsars}
\label{tab:kinematics}
\begin{center}
\begin{tabular}{rrrrrrrr}
\tableline
\\
$\sigma_V$ \hfil & $\sigma_z$ & $\overline{v}_R$ & $\overline{\delta v}_R$ 
& $\overline{v}_t$ & $\overline{\delta v}_t$ 
& $\overline{v}_z$ & $\overline{\delta v}_z$
 \\
(km s$^{-1}$) & (kpc) \hfil \\
\\
\tableline
\tableline
\multicolumn{6}{l}{Uniform Surface Density (independent of $R$):} \\
\\

 $ 20$ & $ 0.05$ &$   0.02$ & $   1.28$ & $   0.00$ & $   0.86$ & $  -0.00$ & $   0.70$  \\ 
       & $ 0.15$ &$   0.01$ & $   1.28$ & $  -0.00$ & $   0.86$ & $  -0.00$ & $   0.79$  \\ 
 $ 40$ & $ 0.05$ &$   0.00$ & $   1.26$ & $  -0.11$ & $   0.80$ & $  -0.00$ & $   0.66$  \\ 
       & $ 0.15$ &$   0.01$ & $   1.26$ & $  -0.11$ & $   0.80$ & $  -0.01$ & $   0.69$  \\ 
 $ 60$ & $ 0.05$ &$   0.00$ & $   1.18$ & $  -0.22$ & $   0.77$ & $   0.00$ & $   0.64$  \\ 
       & $ 0.15$ &$   0.00$ & $   1.18$ & $  -0.22$ & $   0.77$ & $  -0.00$ & $   0.65$  \\ 
 $ 80$ & $ 0.05$ &$   0.01$ & $   1.09$ & $  -0.29$ & $   0.75$ & $   0.00$ & $   0.61$  \\ 
       & $ 0.15$ &$   0.00$ & $   1.09$ & $  -0.30$ & $   0.75$ & $   0.00$ & $   0.61$  \\ 
 $100$ & $ 0.05$ &$   0.01$ & $   1.01$ & $  -0.35$ & $   0.74$ & $   0.00$ & $   0.58$  \\ 
       & $ 0.15$ &$   0.00$ & $   1.00$ & $  -0.35$ & $   0.74$ & $   0.00$ & $   0.58$  \\ 
\\
\tableline
\multicolumn{6}{l}{Exponential Surface Density (in $R$):} \\
\\

 $ 20$ & $ 0.05$ &$   0.03$ & $   1.24$ & $  -0.17$ & $   0.83$ & $  -0.01$ & $   0.69$   \\ 
       & $ 0.15$ &$   0.01$ & $   1.23$ & $  -0.17$ & $   0.83$ & $  -0.01$ & $   0.79$   \\ 
 $ 40$ & $ 0.05$ &$  -0.01$ & $   1.17$ & $  -0.31$ & $   0.78$ & $  -0.00$ & $   0.63$   \\ 
       & $ 0.15$ &$   0.01$ & $   1.17$ & $  -0.31$ & $   0.78$ & $  -0.00$ & $   0.66$   \\ 
 $ 60$ & $ 0.05$ &$  -0.01$ & $   1.09$ & $  -0.41$ & $   0.73$ & $  -0.00$ & $   0.58$   \\ 
       & $ 0.15$ &$   0.00$ & $   1.09$ & $  -0.41$ & $   0.74$ & $   0.00$ & $   0.60$   \\ 
 $ 80$ & $ 0.05$ &$   0.00$ & $   1.03$ & $  -0.47$ & $   0.69$ & $  -0.00$ & $   0.55$   \\ 
       & $ 0.15$ &$   0.01$ & $   1.02$ & $  -0.47$ & $   0.70$ & $   0.00$ & $   0.56$   \\ 
 $100$ & $ 0.05$ &$   0.01$ & $   0.97$ & $  -0.50$ & $   0.66$ & $  -0.00$ & $   0.51$   \\ 
       & $ 0.15$ &$   0.00$ & $   0.96$ & $  -0.50$ & $   0.66$ & $   0.00$ & $   0.52$   \\ 
\\
\tableline
\end{tabular}
\end{center}
Velocity moments for particles of Galactocentric radius $7.5 < R < 9.5$ kpc 
and $|z|<3$ kpc. Here $\overline{v}_R$ means $<v_R>/\sigma_V$, $\overline{\delta
v}_R$ means $\sqrt{<(v_R-<v_R>)^2>}/\sigma_V$, and so on. The top
section refers to a disk with constant birth density in $R$; the bottom
section to a disk with birth density varying with exponential scale
length $3.5$ kpc in $R$.  All moments given in units of $\sigma_V$. Numerical
accuracy of $\pm 0.02$ for all entries. All mixed second order moments are
zero to $\pm 0.03$.
\end{table}

\clearpage

\end{document}